\documentclass[appendixRoman]{ieeeaccess}

\usepackage{cite}
\usepackage{amsmath,amssymb,amsfonts}
\usepackage{algorithmic}
\usepackage{graphicx}
\usepackage{textcomp}
\usepackage{xcolor}
\usepackage{marvosym}
\usepackage{amsthm}
\usepackage{physics}
\usepackage{enumitem}
\usepackage{mathtools}

\newtheorem{theorem}{Theorem}

\newtheorem{lemma}{Lemma}

\newcommand{\pddots}{\vphantom{\ddots}}

\DeclareMathOperator{\adj}{\mathrm{adj}}

\def\BibTeX{{\rm B\kern-.05em{\sc i\kern-.025em b}\kern-.08em
   T\kern-.1667em\lower.7ex\hbox{E}\kern-.125emX}}

\begin{document}
\history{}
\doi{}
\title{Tools for the analysis of quantum protocols requiring state generation within a time window} 

\author{\uppercase{Bethany Davies}\authorrefmark{1,2,3}, \uppercase{Thomas Beauchamp}\authorrefmark{1,2,3}
\uppercase{Gayane Vardoyan\authorrefmark{1,3}, and Stephanie Wehner}.\authorrefmark{1,2,3}}
\address[1]{QuTech, Delft University of Technology, Lorentzweg 1, 2628 CJ Delft, The Netherlands}
\address[2]{Kavli Institute of Nanoscience, Delft University of Technology, Lorentzweg 1, 2628 CJ Delft, The Netherlands }
\address[3]{Quantum Computer Science, EEMCS, Delft University of Technology, Lorentzweg 1, 2628 CJ Delft, The Netherlands}

\corresp{Corresponding authors: Bethany Davies (email: b.j.davies@tudelft.nl), Stephanie Wehner (email: s.d.c.wehner@tudelft.nl).}

\begin{abstract}
Quantum protocols commonly require a certain number of quantum resource states to be available simultaneously. An important class of examples is quantum network protocols that require a certain number of entangled pairs. Here, we consider a setting in which a process generates a quantum resource state with some probability $p$ in each time step, and stores it in a quantum memory that is subject to time-dependent noise. To maintain sufficient quality for an application, each resource state is discarded from the memory after $w$ time steps. Let $s$ be the number of desired resource states required by a protocol. We characterise the probability distribution $X_{(w,s)}$ of the ages of the quantum resource states, once $s$ states have been generated in a window $w$. Combined with a time-dependent noise model, the knowledge of this distribution allows for the calculation of fidelity statistics of the $s$ quantum resources. We also give exact solutions for the first and second moments of the waiting time $\tau_{(w,s)}$ until $s$ resources are produced within a window $w$, which provides information about the rate of the protocol. Since it is difficult to obtain general closed-form expressions for statistical quantities describing the expected waiting time $\mathbb{E}(\tau_{(w,s)})$ and the distribution $X_{(w,s)}$, we present two novel results that aid their computation in certain parameter regimes. The methods presented in this work can be used to analyse and optimise the execution of quantum protocols. Specifically, with an example of a Blind Quantum Computing (BQC) protocol, we illustrate how they may be used to infer $w$ and $p$ to optimise the rate of successful protocol execution.
\end{abstract}

\begin{keywords}
Quantum networks, performance analysis, scan statistics
\end{keywords}

\titlepgskip=-15pt

\maketitle

\section{Introduction}
\label{sec:introduction}
It is common for quantum computing and quantum network protocols to require the simultaneous availability of a certain number of high quality quantum resource states. In the domain of quantum networks, such resource states are typically entangled pairs of qubits, where the execution of protocols such as entanglement distillation and many quantum network applications require multiple entangled pairs to be available at the same time \cite{bennett1996purification,deutsch1996quantum,wehner2018vision}. Another example of a resource state can be found in the domain of quantum computing, where magic state distillation relies on the presence of multiple initial magic states  \cite{bravyi2005universal}.

Here, we study the setting in which resource states are generated using a probabilistic process. In each time step, this process succeeds in generating one resource state with probability $p$. If the state is prepared successfully, it is immediately stored in a quantum memory that is subject to time-dependent noise. The process is repeated until all $s$ states required by a protocol are in memory. Such a setting is ubiquitous in quantum networking, and (photonic) quantum computing. A prime example is heralded entanglement generation~\cite{barrett2005efficient, cabrillo1999creation} that is commonly used in present-day quantum networks (see e.g. \cite{baier2021realization,entanglement_delivery}). 

If the noise is time-dependent, this means that when a state is placed in a quantum memory its quality will degrade over time. In practice then, in order to deliver states of sufficient quality, one often imposes a window of $w$ time steps within which all $s$ states must be produced. If the states are produced within the desired window, the quality of the states is high enough for the application to succeed. Otherwise, the states are typically discarded (see Figure \ref{fig:seq_gen}). In the context of quantum repeater protocols, such a window size is also often referred to as a `cut-off time', and the analysis across multiple nodes is generally non-trivial~\cite{li2020efficient, policies_elementary_links, praxmeyer2013reposition, shchukin2019waiting, optimal_entanglement_dist_policies, rozpkedek2018parameter}. If a protocol requires $s$ quantum resource states of sufficiently high quality to exist simultaneously, this translates to a requirement of $s$ successful generation events within the window of $w$ time steps.

When analysing the performance of protocols that rely on such a generation of resource states, we are interested in a number of performance metrics. For example, one may be interested in the rate at which we can execute a protocol, the probability that the overall quantum protocol will be successful, or a combined metric that considers the number of successful executions of the quantum protocol per time unit. To understand and optimise such performance metrics, we are interested in understanding a number of quantities related to the system's ability to prepare the resource states required by the protocol.

Firstly, one may consider the \textit{waiting time} $\tau_{(w,s)}$ until there are $s$ successes within a window of $w$ time steps. We remark that for fixed parameters ($w,s,p$) this provides us also with information about the rate at which a protocol can be carried out, when executed multiple times.  Secondly, we look at the \textit{ending pattern} $X_{(w,s)}$ (see Figure~\ref{fig:ending_patterns}), which contains the ages of the $s$ quantum resources at time $\tau_{(w,s)}$. Combined with a model of decoherence, this can be used to compute the quality (fidelity) of the resources immediately after the last state has been produced, which is when the quantum protocol may be executed.

The goal of this work is to provide tools that may be used to analyse the performance of a given quantum protocol for specific choices of $w$, $s$ and $p$, as well as to choose a combination of these parameters to optimise its performance.

\Figure[t!](topskip=0pt, botskip=0pt, midskip=0pt)[width=85mm]{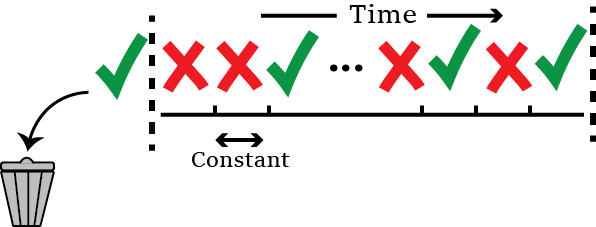}
{\textbf{Setup.} In each time step, a probabilistic process generates a resource state, where $p$ is the probability of success (tick) and $1-p$ the probability of failure (cross). After generation the resource state is immediately placed into a quantum memory subject to time-dependent noise. To ensure the states have sufficient quality to enable a quantum protocol, states that are older than a specific window $w$ of time steps are discarded (bin). \label{fig:seq_gen}}

\Figure[t!](topskip=0pt, botskip=0pt, midskip=0pt)[width=60mm]{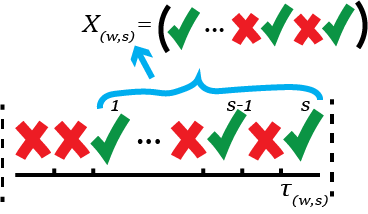}
{\textbf{Ending patterns.} At the first instance $\tau_{(w,s)}$ when a window of $w$ time steps contains $s$ successes, we are interested in how long ago each of the $s$ links were generated. This allows one to quantify the quality of the corresponding resource states. The information of when the $s$ resource states were produced is contained in the ending pattern $X_{(w,s)}$. \label{fig:ending_patterns}}

\subsection{Results}
Our main contributions are summarised below.
\begin{itemize}
    \item For all values of $w$, $s$ and $p$, we provide formulae for the first and second moments of $\tau_{(w,s)}$, and the full distribution of $X_{(w,s)}$. For $(w,s)=(\infty,s)$, these are in a simple closed form, and similarly for $(w,s)=(w,2)$. For all other values of $(w,s)$ we present general formulae, which are in the form of a linear system that may be solved numerically. The dimension of the system scales as $w^{s-1}$. For large $w$ and $s$, it is therefore difficult to obtain closed-form expressions from these systems. 
    \item We provide an efficient method to find bounds on the range of $w$ and $p$ for which $\mathbb{E}(\tau_{(w,s)})$ and $X_{(w,s)}$ may be approximated by $\mathbb{E}(\tau_{(\infty,s)})$ and $X_{(\infty,s)}$ to an arbitrary degree of accuracy. In a practical context, this allows one to quickly compute thresholds on the window size such that increasing $w$ further provides no improvement for a protocol rate. Moreover, for appropriate parameter regimes, this approximation is desirable due to the fact that the dimension of the linear system to solve for the expected waiting time $\mathbb{E}(\tau_{(w,s)})$ and $X_{(w,s)}$ scales with $w$ and $s$, as described above. This is in contrast to the simple closed-from expressions that can be found for the corresponding quantities in the case $w=\infty$. 
    \item We characterise the behaviour of $\mathbb{E}(\tau_{(w,s)})$ and $X_{(w,s)}$ in the limit of a small probability of success. In particular, we show that as $p\rightarrow 0$, $\mathbb{E}(\tau_{(w,s)})$ scales as $p^{-s}$, and that the distribution of $X_{(w,s)}$ becomes uniform. This result may be used to gain intuition about the performance of a quantum application when the resource generation success probability is small, without needing to perform (potentially lengthy) numerical computations.
    \item We provide a demonstration of how these methods may be used in practice. We consider a Blind Quantum Computation protocol \cite{vbqc}. In our model, entanglement is consumed in the transmission of qubits from a client device to a server device. The model includes noise due to imperfect entangled links and memory decoherence. For a set-up involving a computation on four qubits, we provide an example of how the methods from the first sections may be used to choose architecture parameters that optimise the rate of the protocol.
\end{itemize}

\subsection{Related work}
To obtain our results, we draw on methods used in the mathematical literature known as scan statistics. Scan statistics is typically concerned with patterns and clusters in a sequence of random events. This is a field that incorporates techniques from multiple different areas of mathematics. In the quantum context, the problem is different from other areas in caring about the ending pattern distribution. In this work, we therefore use the approach that makes use of martingales, because this allows one to obtain both $\mathbb{E}(\tau_{(w,s)})$ and the distribution of $X_{(w,s)}$ \cite{li_martingale_1980}. It is possible to obtain the same quantities with embedded Markov chains \cite{fu2003distribution}, but we continue here with the martingale method; firstly because the resulting formula has a smaller dimension, and secondly because it has a regular structure that allows us to derive asymptotic results for small $p$, which is an experimentally relevant regime. We note that there exist other methods to obtain $\mathbb{E}(\tau_{(w,s)})$, and also in principle the full distribution of $\tau_{(w,s)}$ - see e.g. \cite{ebneshahrashoob2005sequential} or \cite{scan_statistics_methods_apps} which give formulae to obtain the probability generating function of $\tau_{(w,s)}$. To our knowledge, however, these also result in a large system of equations, and moreover do not give the ending pattern distribution. Other closely related quantities to $\tau_{(w,s)}$ have been explored in great depth in the scan statistics literature, which may also have relevance to the quantum domain. For example, there exist a number of bounds and approximations for $\mathbb{P}(\tau_{(w,s)}\leq n)$  (see e.g. \cite{glaz2001scan} for an overview of results), which may prove useful in allocating time for entanglement generation in a quantum network schedule. By contrast, in this work, we focus on the behaviour of $\mathbb{E}(\tau_{(w,s)})$ and $X_{(w,s)}$, and their implications for the performance of quantum protocols. To our knowledge, this work is the first to characterise the behaviour of the ending pattern distribution in certain parameter regimes, and demonstrate an explicit example of the application of results from scan statistics to a quantum protocol.

\subsection{Outline}
The rest of the paper is organised as follows. In Section \ref{sec:preliminaries}, the quantities $\tau_{(w,s)}$ and $X_{(w,s)}$ are formally defined. In Sections \ref{sec:infinite_window} and \ref{sec:solutions_finite_window}, we give formulae for the first and second moments of $\tau_{(w,s)}$, and the distribution of $X_{(w,s)}$. In Sections \ref{sec:infinite_approximaton} and \ref{sec:asymptotic_behaviour}, we present results that aid the understanding and approximation of these quantities. In Section \ref{sec:analysis}, the behaviour and practical relevance of the results of Section \ref{sec:solutions} are outlined, specifically looking at $\mathbb{E}(\tau_{(w,s)})$. In Section \ref{sec:bqc}, we give an example of how one may use the results of Section \ref{sec:solutions} to choose architecture parameters that optimise the performance of BQC protocol. Finally, further directions are summarised in Section \ref{sec:further_directions}.

\section{Preliminaries}
\label{sec:preliminaries}
We view quantum resource generation attempts as a sequence of i.i.d. Bernoulli trials $(Z_i)_{i = 1}^{\infty}$ with success probability $p = \mathbb{P}(Z_1=1) > 0$. Then, if a protocol requires $s\leq w$ quantum resources to coexist, the time taken to complete the application is dependent on the \textit{waiting time} $\tau_{(w,s)}$ to produce $s$ successes within a window of size $w$. We are also interested in the \textit{ending pattern} $X_{(w,s)}$ which completes the process, because this contains the ages of the $s$ quantum resources present at time $\tau_{(w,s)}$. We denote the set of possible ending patterns as $\Omega(w,s)$. This contains every possible configuration of the $s$ successes within the scanning window, so that $X_{(w,s)}\in \Omega{(w,s)}$. A visualisation of how an ending pattern realises the end of the process is given in Figure \ref{fig:ending_patterns}. More specifically, we define
\begin{equation}
    \Omega_l(s) := \{x \in \{0,1\}^l: x_1 = x_l = 1~ \wedge~ \sum_{i = 1}^l x_i = s \}
    \label{eqn:ending_patterns_def}
\end{equation}
to be the set of all length-$l$ binary strings  $x = (x_1,...,x_l)$ that contain $s$ successes, two of which occur at either end of the string. Then,
\begin{equation}
    \Omega(w,s) := \bigcup_{l = s}^w \Omega_l(s)
    \label{eqn:Omega_def}
\end{equation}
is the set of ending patterns.  The set $\Omega(w,s)$ can be thought of as containing all clusters of $s$ successes that were produced within a time less than or equal to $w$ time steps. Note that the number of possible ending patterns is given by
\begin{equation}
  |\Omega(w,s)| = {w-1 \choose s-1}. 
\label{eqn:size_omega}
\end{equation}
To see this, consider the fact that each ending pattern in $\Omega(w,s)$ corresponds uniquely to an ending scenario where the $s$ successes are distributed within the window of $w$ time steps. Since the final quantum resource must always have been prepared at the most recent time step and therefore is fixed, it remains to distribute the remaining $s-1$ successes within $w-1$ time steps, meaning that the number of possible ending patterns is restricted to (\ref{eqn:size_omega}). The waiting time $\tau_{(w,s)}$ is then defined by
\begin{equation}
   \tau_{(w,s)} := \min\limits_{x \in \Omega(w,s)}\{\tau_x \}
   \label{eqn:waiting_time_compound_def},
\end{equation}
i.e. this is the time until we see the first ending pattern in the sequence of Bernoulli trials. Here, $\tau_x$ is the time taken until one particular ending pattern $x$ is first seen, so that for $x \in \Omega_l(s) \subset \Omega(w,s)$ 
\begin{equation}
    \tau_x := \text{min}\{t: (Z_{t-l+1},Z_{t-l+2},...,Z_{t}) = x  \}.
    \label{eqn:wt_simple_def}
\end{equation}
We note that $\tau_{(w,s)}$ is well-defined because it is bounded above by a geometric random variable (see Appendix \ref{sec:waiting_time_well_defined}).
There is also a relationship between $\tau_{(w,s)}$ and the distribution of $X_{(w,s)}$ given by
\begin{equation}
    \mathbb{P}(X_{(w,s)}=x) = \mathbb{P}(\tau_{(w,s)} = \tau_x),   \label{eqn:ending_pattern_dist_def}
\end{equation}
recalling that $X_{(w,s)}$ takes the value of the ending pattern that completes the process. No two ending patterns can realise the end of the process at the same time since no element of $\Omega(w,s)$ contains another, and so $X_{(w,s)}$ is well-defined.

\section{Formulae and approximations}
\label{sec:solutions_approximations}
In the following two sections, we provide exact solutions for the first and second moments of $\tau_{(w,s)}$, and the full distribution of $X_{(w,s)}$. Formulae are provided for all possible values of $w$ and $s$. In Section \ref{sec:infinite_approximaton}, we look at approximating the solutions for a large $w$. In Section \ref{sec:asymptotic_behaviour}, we characterise the solution behaviour for small $p$.
\label{sec:solutions}
\subsection{Infinite window}
\label{sec:infinite_window}
Here, we consider the case where no resource states are discarded (or equivalently when $w=\infty$) and give solutions for the first and second moments of $\tau_{(\infty,s)}$, and the distribution of $X_{(\infty,s)}$. This serves as a useful initial study of the problem, providing intuition for the case where $w$ is large and finite.

When no states are discarded, the waiting time to see all of the successes simply becomes a sum of $s$ i.i.d. geometric random variables with parameter $p$. This is known as a \textit{negative binomial} distribution, and has an exact distribution given by 
\begin{equation}
    \mathbb{P}(\tau_{(\infty,s)} = n) =  {n-1 \choose s-1}(1-p)^{n - s}p^{s}
    \label{eqn:wt_dist_w=infty}
\end{equation}
and expectation 
\begin{equation}
    \mathbb{E}(\tau_{(\infty,s)}) = \frac{s}{p}.
\label{eqn:expectation_infinite}
\end{equation}
Note that for $w'>w$, it is always the case that $\tau_{(w',s)}\leq \tau_{(w,s)}$, and so
\begin{equation}
    \mathbb{E}(\tau_{(w',s)})\leq \mathbb{E}(\tau_{(w,s)}), \text{ for }w'>w.
\label{eqn:waiting_time_decreasing_fn_of_w}
\end{equation}
In particular, the waiting time for a finite $w$ will always be greater than or equal to the infinite case. Then, (\ref{eqn:expectation_infinite}) gives a simple lower bound in terms of $s$ and $p$
\begin{equation}
\mathbb{E}(\tau_{(w,s)}) \geq \frac{s}{p}.
\label{eqn:lower_bound}
\end{equation}
The variance of $\tau_{(\infty,s)}$ is given by 
\begin{equation}
    \text{Var}(\tau_{(\infty,s)}) = \frac{s\cdot (1-p)}{p^2},
    \label{eqn:variance_infinite_window}
\end{equation}
from which we can see that the standard deviation is reciprocal in $p$. 

It is also possible to derive a simple expression for the distribution of $X_{(\infty,s)}$. For a binary string $x \in \Omega_l$ that lives in the (now infinite) set of ending patterns $\Omega(\infty,s)$, 
\begin{equation}
    \mathbb{P}(X_{(\infty,s)} = x) = (1-p)^{l-s}p^{s-1},
    \label{eqn:end_cluster_prob_inf}
\end{equation}
which can be seen by considering the probability of generating the remaining $l-1$ entries of $B$ after the first success has been generated. We see that when the window size is infinite, the probability of generating ending patterns of the same length is constant.

\subsection{Finite window}
\label{sec:solutions_finite_window}
\subsubsection{$s=2$}
\label{sec:s=2}
When $s=2$, it is possible again to derive closed-form solutions for the first and second moment of $\tau_{(w,s)}$ and the distribution of $X_{(w,s)}$. We present below the formulae for $\mathbb{E}(\tau_{(w,s)})$ and the ending pattern distribution.

In this case, the ending patterns are determined by the time between the two states, i.e. $|\Omega_l(2)|=1$. We separate the process of resource generation into two parts: generation of the first state, and generation of the second state. Generation of the first link occurs when there is no state stored in memory. This is not limited by the window, and has a generation time described by a geometric distribution with parameter $p$. To finish the process, the generation of the second state must happen within $w-1$ time steps of the first link being generated. When the process is finished, the time between the two states is then a geometric distribution conditional on this event, which occurs with probability $1-(1-p)^{w-1}$. Then, letting $L\in \{ 1,...,w-1\}$ be the number of attempts after the first state to generate the second,
\begin{equation}
    \mathbb{P}(L=n) = \frac{(1-p)^{n-1}p}{1-(1-p)^{w-1}},
    \label{eqn:ending_pattern_dist_s=2}
\end{equation}
which gives the full ending pattern distribution, where $L=n $ corresponds to $X_{(w,s)}\in \Omega_{n+1}(2)$.

Now, let $M$ be the number of times a first state must be generated until the process is finished. Since this is determined by the success of the second state within the time window, we have 
\begin{equation}
    M \sim \text{Geom}\left(1-(1-p)^{w-1}\right).
    \label{eqn:M_dist}
\end{equation}
The total time is then given in terms of $M$ and $L$ by 
\begin{equation}
    \tau_{(w,2)} = \sum_{j = 1}^M T_j  + (M-1)(w-1) + L, 
    \label{eqn:s=2_decomposition}
\end{equation}
where the random variables $T_j \sim \text{Geom}(p)$ describe the number of attempts to generate the first state. Now, as shown in Appendix \ref{sec:appendix_two_links}, 
\begin{equation}
    \mathbb{E}\left( \sum_{j = 1}^M T_j \right)= \mathbb{E}(M)\mathbb{E}(T_1) = \frac{1}{\left(1-(1-p)^{w-1} \right)}\cdot \frac{1}{p},
\end{equation}
and
\begin{equation}
    \mathbb{E}(L)
     = \frac{1-(1-p)^{w}-wp(1-p)^{w-1}}{p(1-(1-p)^{w-1})}.
\end{equation}
The expected waiting time may then be computed as
\begin{gather}
    \mathbb{E}(\tau_{(w,2)}) = \mathbb{E}(M)\mathbb{E}(T_1) + (\mathbb{E}(M)-1)(w-1) + \mathbb{E}(L),
\end{gather}
from which we obtain 
\begin{equation}
     \mathbb{E}(\tau_{(w,2)}) = \frac{1}{p} + \frac{1}{p(1-(1-p)^{w-1})}.
\end{equation}
The variance of $\tau_{(w,2)}$ may also be computed by making use of (\ref{eqn:s=2_decomposition}). The computation is given in Appendix \ref{sec:appendix_two_links}. 
\subsubsection{$s>2$}
We now give a formula to exactly compute $\mathbb{E}(\tau_{(w,s)})$ and the full distribution $\{ \mathbb{P}(X_{(w,s)}=x): x \in \Omega(w,s) \}$, for a finite $w$. This is derived using the method from \cite{li_martingale_1980}, which makes use of martingales. For completeness, we include an outline of the derivation in Appendix \ref{sec:derivation_finite_window_size}, where a gambling analogy is introduced to aid understanding. The resulting formula is in the form of a linear system of size $|\Omega(w,s)|+1$ that can be solved exactly. Each matrix element defining the linear system can be computed simply and efficiently. Further, in Appendix \ref{sec:second_moment} we give a formula for the second moment of $\tau_{(w,s)}$, which now involves two linear systems of size $|\Omega(w,s)|$. A martingale method is also used for its derivation, and for this we refer to \cite{pozdnyakov_martingale_2009}. The second moment of $\tau_{(w,s)}$ can then be used to calculate the variance and standard deviation of $\tau_{(w,s)}$.

Before stating the first formula, we introduce some notation. We define a function $*$ that maps two binary strings $x = (x_1,...,x_k)$ and $y = (y_1,...,y_m)$ to a scalar value, given by 
\begin{equation}
    x*y := \sum_{j = 1}^{\text{min}(k,m)} \left( \prod_{i = 1}^{j} \delta_{(x_i,y_{m-j+i})} \right),
    \label{eqn:star_def}
\end{equation}
where for $a,b \in \{0,1\}$, the quantity $\delta_{a,b}$ is given by 
\begin{equation}
    \delta_{(a,b)} =
    \begin{cases}  \frac{1}{p_a} \text{ if } a = b; \\
     0, \text{ otherwise,}
    \end{cases}
    \label{eqn:delta_definition}
\end{equation}
where $p_1:=p$ and $p_0:=1-p$. From (\ref{eqn:star_def}), we see that the value of $x*y$ is obtained by comparing the overlap of successive substrings of $x$ and $y$. If two substrings match exactly, then the corresponding term is included in the sum, and it is weighted by an amount that is dependent on the Bernoulli parameter $p$. Informally, then, $x*y$ measures how similar the structures of $x$ and $y$ are. A simple example of the action of $*$ is given as follows. We consider the action of $*$ on two ending patterns, recalling (\ref{eqn:ending_patterns_def}). Letting $s=3$ and $w\geq 7$, suppose that $x = (1,0,1,0,0,0,1)$ and $y = (1,0,0,0,1,1)$. Computing (\ref{eqn:star_def}) then yields
\begin{equation*}
    x*y = \frac{1}{p_1} + 0 + 0 + 0 +  \frac{1}{p_1^2}\cdot \frac{1}{p_0^3} + 0 = \frac{1}{p_1} + \frac{1}{p^2 p_0^3}.
\end{equation*}
Since all elements of $\Omega(w,s)$ start and finish with a success, the same initial $1/p$ term will be present for any pair of ending patterns. Whether or not there are higher order terms depends on the overlap of the successive substrings. In particular, for two ending patterns $x,y \in \Omega(w,s)$, the quantity $x*y$ will be of order $1/p^s$ if and only if $x=y$. 

Equipped with these definitions, we now give the formula for the expected waiting time and the ending pattern distribution. 
\begin{theorem}
   Let $N:=|\Omega(w,s)|$. After enumerating the ending patterns as $\Omega(w,s) \equiv \{x^{(i)} : i = 1,..., N  \}$, let
\begin{equation}
    \vec{v} =  \left(
 \begin{array}{c}
 \mathbb{E}(\tau_{(w,s)}) \\
 \mathbb{P}\left( X = x^{(1)}\right) \\ 
 \mathbb{P}\left( X = x^{(2)}\right) \\
 \vdots\\
 \vdots \\
 \mathbb{P}\left( X=x^{(N)}\right)
 \end{array}
 \right).
\end{equation}
Then 
\begin{equation}
    A\vec{v} = \vec{e_1},
    \label{eqn:linear_system}
\end{equation}
where $\vec{e_1}:= (1,0,...,0)^T$ is a vector of length $N+1$, and
\begin{equation}
    A :=    \left( 
    \arraycolsep=2.5pt
     \def\arraystretch{1.5}
    \begin{array}{c c c c c} 
 0 & 1 & 1 & \cdots & 1 \\
 -1 & x^{(1)}*x^{(1)} & x^{(1)}*x^{(2)} & \cdots & x^{(1)}*x^{(N)} \\
 -1 & x^{(2)}*x^{(1)} & x^{(2)}*x^{(1)}  & \cdots & x^{(2)}*x^{(N)} \\
 \vdots & & \ddots & & \vdots \\
 \vdots & &  &\ddots  & \vdots \\
 -1 & x^{(N)}*x^{(1)} & \cdots & & x^{(N)}*x^{(N)}
 \end{array} 
 \right).
 \label{eqn:matrix_definition}
\end{equation}
\label{thm:exact_solutions}
\end{theorem}
The matrix $A$ is invertible since no element of $\Omega(w,s)$ contains another \cite{li_martingale_1980}. The solution for $\vec{v}$ therefore always exists and is unique. The matrix $A$ is fully determined by the success probability $p$ and the parameters $w$ and $s$. As discussed above, each entry of the submatrix formed by removing the first row and column is greater than or equal to $1/p$. Entries that do not take this exact value contain higher-order terms in $1/p$, due to the fact that there is a greater overlap of the ending patterns corresponding to the row and column indices of such an entry. The entries of the highest power in $1/p$ are exactly the diagonal elements and are of order $s$, because a string overlaps completely with itself and contains $s$ successes. In principle, the solutions for $\vec{v}$ can be computed analytically as functions of $p$ by inverting $A$ directly. Noting that the dimension of the system is $|\Omega(w,s)|+1 = \mathcal{O}(w^{s-1}) $, this is in practice computationally laborious for large $w$ and $s$.  In the following sections, we derive results that aid the understanding and computation of the corresponding results in the two characteristic regimes of large $w$, and small $p$.

\subsection{Approximating with an infinite window}
\label{sec:infinite_approximaton}
Now, one might ask: how large must the window size be for the approximation $w = \infty$ to be accurate? This is desirable due to the simple analytical form of the results for the distributions of $\tau_{(\infty,s)}$ and $X_{(\infty,s)}$, as seen in Section \ref{sec:infinite_window}. This is in contrast to the solutions presented in Section \ref{sec:solutions_finite_window} for the case of a finite $w$, which scale with $w$ and $s$. The approximation becomes valid when the window size has `saturated' the process, so that increasing the window size does not provide any significant improvement for the rate. Alternatively, the approximation becomes accurate when $\mathbb{P}(\tau_{(w,s)}>w)$ is small. This intuition is formalised with the following theorem. 
\begin{theorem}
    Let $\tau_{(w,s)}$ be the waiting time for $s$ successes in a $w$-window. Let $X_{(w,s)}$ be the corresponding ending pattern. Let $p$ denote the success probability of each trial. Let $\epsilon(w,s,p)\coloneqq \mathbb{P}(\tau_{(w,s)}>w)$. Suppose that  $0<p<1$ and $w<\infty$. Then
    \begin{equation}
    \frac{\mathbb{E}(\tau_{(w,s)})-\mathbb{E}(\tau_{(\infty,s)})}{\mathbb{E}(\tau_{(w,s)})} < \epsilon(w,s,p)
    \label{eqn:difference_bound}
    \end{equation}
    and 
    \begin{equation}
    \sum_{x\in \Omega(\infty,s)}\! |\mathbb{P}(X_{(w,s)}\!=\!x)-\mathbb{P}(X_{(\infty,s)}\!=\!x)| < 2\epsilon(w,s,p).
    \end{equation} 
    \label{thm:infinite_w_bound}
\end{theorem}
We now look to evaluate $\epsilon(w,s,p)$. Looking back at the identity (\ref{eqn:wt_dist_w=infty}) for the distribution of $\tau_{(\infty,s)}$, we have 
\begin{gather}
    \mathbb{P}(\tau_{(w,s)}> w) = \mathbb{P}(\tau_{(\infty,s)}> w) \\ = \sum_{n = w+1}^{\infty} {n-1 \choose s-1}(1-p)^{n - s}p^{s}.
\end{gather}
To evaluate the right-hand side of (\ref{eqn:difference_bound}) it is convenient to rewrite this as a finite sum, as provided by the following lemma. The proof of this is given in Appendix \ref{sec:appendix_sum_evaluation}.
\begin{lemma}
   Let $\tau_{(w,s)}$ be the waiting time for $s$ successes in a $w$-window, as defined in (\ref{eqn:waiting_time_compound_def}). Suppose that  $0<p<1$ and $w<\infty$. Then
\begin{equation}
    \mathbb{P}(\tau_{(w,s)}> w) = \sum_{i=0}^{s-1} {w \choose i}(1-p)^{w - i}p^{i}.
    \label{eqn:epsilon_identity}
\end{equation}
\label{lem:epsilon_sum}
\end{lemma}
This sum is simple and efficient to evaluate for constant $s$, and can then be used to find a range of $w$ for which the two expectations are close. For example, if one is interested in evaluating $\mathbb{E}(\tau_{(w,s)})$, when in fact $w$ is large enough such that it may be reasonably approximated by $\mathbb{E}(\tau_{(\infty,s)}) = s/p$, it is possible avoid solving a large system of equations with the following method. Demanding some desired error $\delta$, one may quickly compute 
\begin{equation} 
    w^{*} = \text{min}\big\{ w: \epsilon(w,s,p)< \delta \big\}.
\end{equation}
Then, for all $w\geq w^*$, one may approximate $\mathbb{E}(\tau_{(w,s)})$ with $\mathbb{E}(\tau_{(\infty,s)}) = s/p$ with accuracy on the order of $\delta$. The same can be done with the ending pattern distribution: if one is interested in the expectation of some fidelity quantity $F(X_{(w,s)})$, one may also approximate $\mathbb{E}(F(X_{(w,s)}))$ with $\mathbb{E}(F(X_{(\infty,s)}))$ with the same accuracy.

\subsection{Asymptotic behaviour of the expectation}
\label{sec:asymptotic_behaviour}
From (\ref{eqn:waiting_time_decreasing_fn_of_w}), an upper bound for $\mathbb{E}(\tau_{(w,s)})$ is given by $\mathbb{E}(\tau_{(s,s)})$, which can be written in a simple analytical form. In the case $w=s$, there is only one ending pattern $x$, which corresponds to the case of $s$ consecutive successes. From (\ref{eqn:linear_system}), we then have
\begin{equation}
    \mathbb{E}(\tau_{(s,s)}) = x*x = \sum_{j = 1}^s \frac{1}{p^j} = \frac{1/p^s - 1}{1-p},
    \label{eqn:exp_wt_w=s}
\end{equation}
which for small $p$ satisfies $\mathbb{E}(\tau_{(s,s)}) \sim 1/p^s$. In comparison, from (\ref{eqn:expectation_infinite}), the scaling of the expectation for $w = \infty $ is reciprocal in $p$. Further, from the form of $A$ given in (\ref{eqn:matrix_definition}), all entries of $\vec{v}$ will be ratios of polynomials in $1/p$. Looking at the first component of $\vec{v}$, which is the waiting time expectation, this tells us that there is some integer value $\alpha_s(w)$ which dominates the scaling for small $p$, so that 
\begin{equation}
    \mathbb{E}(\tau_{(w,s)}) \sim \frac{c(w,s)}{p^{\alpha_s(w)}},
    \label{eqn:small_p_behaviour}
\end{equation}
where $c(w,s)$ is a constant. Now, recalling from (\ref{eqn:waiting_time_decreasing_fn_of_w}) that $\mathbb{E}(\tau_{(w,s)})$ is a decreasing function of $w$, we therefore expect the same of $\alpha_s(w)$, which satisfies $\alpha_s(s) = s$ and $\alpha_s(\infty)= 1$. Below we show that for $w< \infty$, $\alpha_s(w)$ is always equal to $s$, and also derive the scaling factor $c(w,s)$.
\begin{theorem}
   Let $\tau_{(w,s)}$ be the waiting time for $s$ successes in a $w$-window, as defined in (\ref{eqn:waiting_time_compound_def}). Let $X_{(w,s)}$ be the corresponding ending pattern. Let $p$ be the success probability of the process. Then, in the limit $p \rightarrow 0$,
   \begin{equation}
       \mathbb{E}\left( \tau_{(w,s)} \right) \sim \frac{1}{|\Omega(w,s)| p^s},
       \label{eqn:expectation_scaling_const}
   \end{equation}
   and for all $x\in \Omega(w,s)$
   \begin{equation}
       \mathbb{P}(X_{(w,s)}=x) \rightarrow \frac{1}{|\Omega(w,s)|} 
       \label{eqn:ending_pattern_sim_uniform}
   \end{equation}
   where $|\Omega(w,s)|= {w-1 \choose s-1}$ is the number of possible ending patterns.
   \label{thm:asymptotic_small_p}
\end{theorem}
A proof of Theorem \ref{thm:asymptotic_small_p} is given in Appendix \ref{sec:appendix_asymptotics}. It is interesting future work to quantify the speed of convergence of (\ref{eqn:expectation_scaling_const}) and (\ref{eqn:ending_pattern_sim_uniform}). 

As intuition for (\ref{eqn:ending_pattern_sim_uniform}), note that for very small $p$, the probability of having $w$ failures preceding the ending pattern is high. In this case, the ending pattern distribution is equivalent to the ending pattern distribution given we succeed in $w$ attempts, which in the limit of small $p$ converges to the uniform distribution.

The behaviour captured by Theorem \ref{thm:infinite_w_bound} and \ref{thm:asymptotic_small_p} may be viewed as two limiting behaviours of the problem in the regimes of small and large $p$, respectively. In  particular, we expect that the formula provided by Theorem \ref{thm:exact_solutions} becomes useful in neither regime, i.e. when $p$ is neither too small or too large to apply either approximation. Moreover, it is important to keep in mind that such a regime will depend on the choices of $w$ and $s$.

\section{Illustration and application}
\label{sec:analysis}
We expect the methods presented in the above sections to be useful in choosing the optimal window size for a quantum protocol. To this end, we firstly analyse in more detail the behaviour of $\mathbb{E}(\tau_{(w,s)})$. We then demonstrate how these methods may be used to optimise the performance of a BQC protocol.

\subsection{Illustration}
\label{sec:analysis}
We fix $s=4$ as an example to showcase the characteristic behaviours of the expected waiting time. From our investigations, the solutions for other values of $s$ display the same qualitative behaviour. The value $s=4$ is small enough that the complexity of the problem is not too large to be solved on a laptop.

\Figure[t!](topskip=0pt, botskip=0pt, midskip=0pt)[width=85mm]{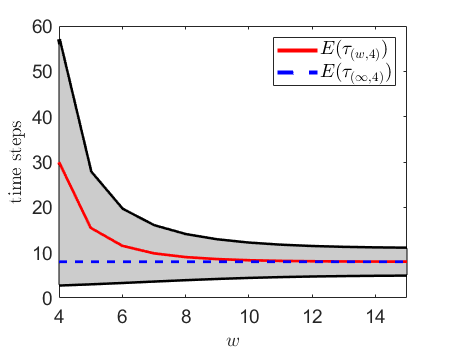}
{\textbf{How $\mathbb{E}(\tau_{(w,4)})$ varies with $w$.} We see that $\mathbb{E}(\tau_{(w,4)})$ (red line) converges to  the lower bound $\mathbb{E}(\tau_{(\infty,4)})=4/p$ (blue line) as $w$ becomes large. The grey region is one standard deviation of $\tau_{(w,4)}$ above and below its expectation (for $w<\infty$). All quantities are evaluated with a success probability $p=0.5$. \label{fig:window_vs_expectation}} 

In Figure \ref{fig:window_vs_expectation},  $\mathbb{E}(\tau_{(w,4)})$ is plotted against $w$, with the success probability set to $p=0.5$. We notice the convergence to the $w=\infty$ lower bound. The grey region is that given by one standard deviation above and below the expectation. Note that one also expects the standard deviation to converge to that of $\tau_{(\infty,4)}$, which is given in closed form by (\ref{eqn:variance_infinite_window}).

\Figure[t!](topskip=0pt, botskip=0pt, midskip=0pt)[width=85mm]{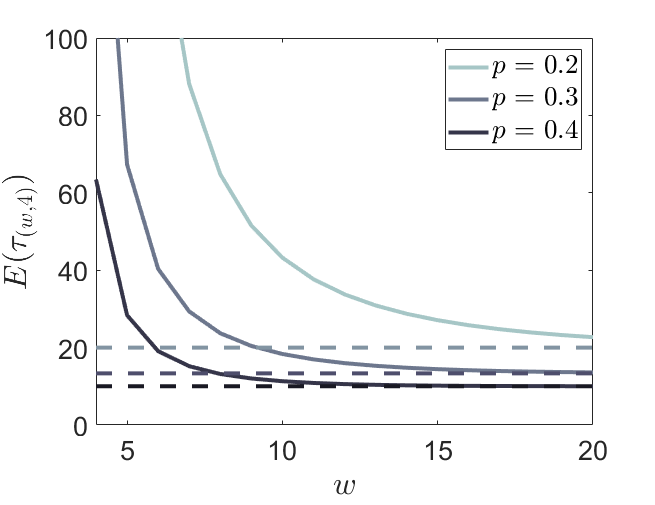}
{\textbf{How $\mathbb{E}(\tau_{(w,4)})$ varies with $w$ and $p$} We see that for larger $p$, $\mathbb{E}(\tau_{(w,4)})$ (solid line) approaches the lower bound $\mathbb{E}(\tau_{(\infty,4)})=4/p$ (dashed line) at a higher rate.
\label{fig:window_vs_expectation_vary_p}}

\Figure[t!](topskip=0pt, botskip=0pt, midskip=0pt)[width=85mm]{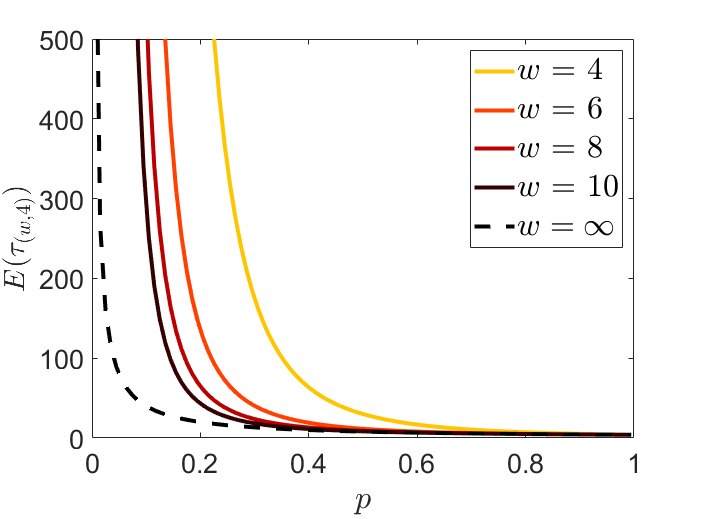}
{\textbf{How $\mathbb{E}(\tau_{(w,4)})$ varies with $p$.} We see that $\mathbb{E}(\tau_{(w,4)})$ (solid lines) demonstrates the reciprocal scaling as $p \rightarrow 0$, as encapsulated by Theorem \ref{thm:asymptotic_small_p}. There is convergence to $\mathbb{E}(\tau_{(w,4)})$ (dashed line). This plot was made by discretising $p$ into 100 points, evenly spaced in the range $(0,1)$. \label{fig:prob_vs_expectation} }

In Figure \ref{fig:window_vs_expectation_vary_p}, $\mathbb{E}(\tau_{(w,4)})$ is again plotted against $w$, but this time for three different values of the success probability. In each case, the solution again approaches the corresponding $w=\infty$ lower bound. Each line starts at $\mathbb{E}(\tau_{(4,4)}) = (1/p^s - 1)/(1-p)$, corresponding to $w=s$, and converges to the $w=\infty$ limit. This convergence is an important feature, because at some point increasing $w$ provides no significant improvement for the protocol rate. As one would expect intuitively, the convergence occurs more quickly for a larger $p$, as increasing the window size effectively saturates the problem more easily. To quantify this, we can use the arguments of Section \ref{sec:infinite_approximaton}. For example, taking the desired margin of error to be $2\%$, define
\begin{equation}
    w^{*} = \text{min}\left\{ w: \epsilon(w,s,p)<0.02 \right\},
    \label{eqn:w_star_def}
\end{equation}
where $\epsilon(w,s,p)$ is given by (\ref{eqn:epsilon_identity}). By Theorem \ref{thm:infinite_w_bound} and Lemma \ref{lem:epsilon_sum}, the approximation $\mathbb{E}(\tau_{(w,s)}) \approx \mathbb{E}(\tau_{(\infty,s)})$ is then valid to the same margin of error for all $w>w^*$. It is interesting to see how this compares to the smallest window size $w^*_{\text{true}}$ for which the same approximation can be made, which is defined formally as 
\begin{equation}
    w^{*}_{\text{true}} = \text{min}\bigg\{w : \frac{\mathbb{E}(\tau_{(w,s)})-\mathbb{E}(\tau_{(\infty,s)})}{\mathbb{E}(\tau_{(w,s)})} < 0.02 \bigg\}.
    \label{eqn:w_star_true}
\end{equation}
The value $w^*$ is then an upper bound for $w^*_{\text{true}}$. For example, letting $p = 0.5$ and $s=4$ yields $w^{*}=15$, and checking with the exact solutions gives $w^{*}_{\text{true}} = 12$. These are plotted for more values of $p$ in Figure \ref{fig:w_star_ub}. We see from the plot that as $p$ increases, the bound appears to become tighter. The value $w^*_{\text{true}}$ was plotted for only a few select values of $p$ because its calculation is computationally intensive.

 \Figure[t!](topskip=0pt, botskip=0pt, midskip=0pt)[width=85mm]{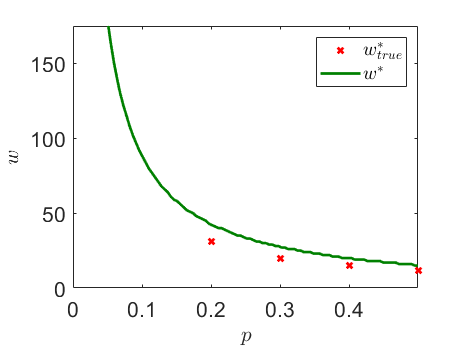}
{\textbf{Comparison of thresholds on $w$ for the infinite window approximation}. One may use $w^{*}$ (green line) as a threshold, which is more easily computable than $w^{*}_{\text{true}}$ (red cross). See (\ref{eqn:w_star_def}) and (\ref{eqn:w_star_true}) for the definition of these quantities. Here we assume a desired error of $2\%$. \label{fig:w_star_ub}}

In Figure \ref{fig:prob_vs_expectation},  $\mathbb{E}(\tau_{(w,4)})$ is plotted against $p$ for five different values of the window size. One can see that the scaling of $\mathbb{E}(\tau_{(w,4)})$ occurs more slowly for a larger $w$. This indicates the reciprocal behaviour as given in (\ref{eqn:expectation_scaling_const}), where $\mathbb{E}(\tau_{(w,4)})\sim 1/|\Omega(w,4)|p^4$, and in the case of a larger $w$ the constant $|\Omega(w,4)|$ suppresses the scaling. As $p\rightarrow 1$, all plots simply converge to $s = 4$, because in this case the process is deterministic. Further, we see again the convergence of the expectation to the infinite window limit. If $p$ becomes large, we expect the problem to `saturate' in the same sense as before, so that $\mathbb{E}(\tau_{(w,s)}) \approx \mathbb{E}(\tau_{(\infty,s)})$. The speed of this convergence can again be quantified using the results of Section \ref{sec:infinite_approximaton}. Demanding the same error of $2\%$, we take the $p^{*}$ that satisfies 
\begin{equation}
    p^{*} = \text{inf}\left\{ p: \epsilon(w,s,p)<0.02 \right\},
    \label{eqn:p_star_def}
\end{equation}
or equivalently, $p^{*}$ is the unique value of $p$ such that $\epsilon(w,s,p^{*}) = 0.02$. The value $p^*$ is an upper bound for the true threshold $p_{\text{true}}^*$, 
\begin{equation}
    p^{*}_{\text{true}} \coloneqq \text{inf}\bigg\{p:\left(\frac{\mathbb{E}(\tau_{(w,s)})-\mathbb{E}(\tau_{(\infty,s)})}{\mathbb{E}(\tau_{(w,s)})} \right)\bigg\rvert_{p}  < 0.02 \bigg\}.
    \label{eqn:p_star_true}
\end{equation}
where we now include dependence of the expectations on the success probability $p$. In Figure \ref{fig:p_star}, $p^{*}$ and $p^{*}_{\text{true}}$ are plotted against $w$. The value $p^*_{\text{true}}$ is computationally intensive to find for large values of $w$, and has therefore only been plotted for selected small values of $w$. The bound $p^*$, however, is efficient to compute. We observe that the bound appears to be tighter for smaller $w$. 

 \Figure[t!](topskip=0pt, botskip=0pt, midskip=0pt)[width=85mm]{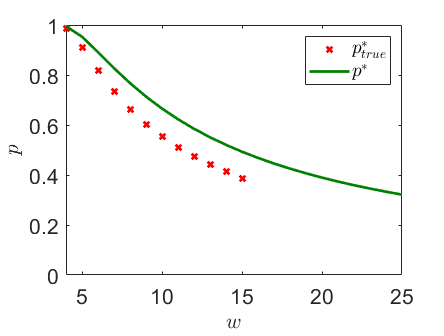}
{\textbf{Comparison of thresholds on $p$ for the infinite window approximation}. One may use $p^{*}$ (green line) as a threshold, which is more easily computable than $p^{*}_{\text{true}}$ (red cross). See (\ref{eqn:p_star_def}) and (\ref{eqn:p_star_true}) for the definition of these quantities. Here we assume a desired error of $2 \%$. \label{fig:p_star}}

\subsection{Application to a BQC protocol}
\label{sec:bqc}
In the following, we provide an example of how the results from Section \ref{sec:solutions} may be used in the performance analysis of a quantum network application. We consider a verifiable Blind Quantum Computation (BQC) protocol \cite{vbqc}. This involves a client, who uses a more powerful server device to carry out a bounded-error quantum polynomial-time (BQP) computation \cite{nielsen_chuang}, which is specified in the measurement-based formalism \cite{one_way_qc}. In this formalism, the computation is defined with respect to a graph $G=(V,E)$, where $V$ is the set of vertices and $E$ the set of edges. The computation is performed by firstly creating a graph state corresponding to $G$, and then applying a series of measurements (`measurement flow') to a subset of qubits. The BQC protocol is designed such that the server remains ignorant of the client's desired computation (blindness). Further, it ensures that the client can validate that the outcome is correct, even in the presence of some amount of noise or a malicious server (veriability). These properties are stated precisely in terms of the composable security properties of the protocol \cite{dunjko2014composable}. For the protocol in full detail, we refer to \cite{vbqc}. Here, we provide a short outline of the BQC protocol, and a simple model of how it is carried out. We then apply the results of Section \ref{sec:solutions_approximations} to study the performance of the protocol.

\subsubsection{Protocol feasibility}
The BQC protocol involves a series of rounds. In each round, the client sends $|V|$ qubits to the server, and also a description of the measurement flow it should carry out. If the server is honest, it will then create a graph state by applying entangling gates corresponding to edges in $E$, carry out the corresponding measurement flow, and send the measurement outcomes back to the client.

The protocol involves interweaving two types of rounds: \textit{computation} and \textit{test} rounds. The computation rounds are used to carry out the client's desired computation. In these rounds, the computation measurement flow is encrypted in order to maintain blindness. The function of the test rounds is to check for deviations from the client's specified operations. Deviations could be due to noise, or the server being malicious. Each test round has the outcome of either pass or fail, and the protocol is aborted if the ratio of failed test rounds lies above a certain threshold. 

Assuming the test round outcomes are i.i.d., the sufficient condition for verifiability that we will consider is given by
\begin{equation}
    p_{\text{av}} < \frac{2\gamma-1}{k(2\gamma-2)},
    \label{eqn:avg_error_condition}
\end{equation}
as shown in \cite{delft_eindhoven}. Here, $p_{\text{av}}$ is the average probability of failure of a test round, and $\gamma$ is the inherent error probability of the BQP computation. The value $k$ is an integer and is corresponding to the $k$-colouring chosen by the client. This is a partition of the set of vertices into $k$ subsets, known as colours, such that there is no edge between two vertices of the same colour. For the relevance of this to the BQC protocol, see Appendix \ref{sec:appendix_test_rounds} for a description of test rounds. For deterministic computations ($\gamma = 0$), (\ref{eqn:avg_error_condition}) simplifies to
\begin{equation}
 p_{\text{av}} < \frac{1}{2k}.
 \label{eqn:avg_error_condition_det}
\end{equation}
When the server is honest, the quantity $p_{\text{av}}$ is determined on the amount by noise, which could for example arise from imperfect local operations and measurements, or imperfect memory in the server. Further, in a networked setting where the client and server are distantly separated, the client may send its qubits to the server by making use of entanglement that has been established between the two parties. In this way, the performance of the protocol is directly dependent on properties of the quantum network architecture connecting client and server. In such architectures, however, there is in general a trade-off between rate and quality. In the case of this BQC protocol, demanding that the condition (\ref{eqn:avg_error_condition}) is met then effectively places an upper bound on the rate of the protocol. In the following, we consider a simple model of the network and device architectures, and provide a demonstration of how the methods presented in Section \ref{sec:solutions} may be used to find architecture parameters that maximise the protocol rate, given the constraint (\ref{eqn:avg_error_condition}).

\subsubsection{Model of network architecture}
\label{sec:bqc_model}
\Figure[t!](topskip=0pt, botskip=0pt, midskip=0pt)[width=85mm]{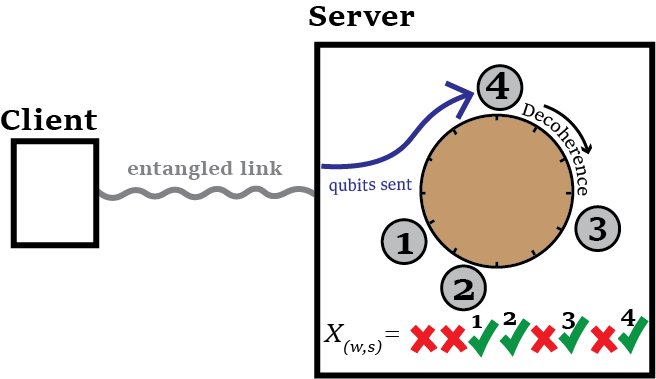}
{\textbf{The scenario considered for the model of BQC.} Generation of the entangled link is attempted sequentially, with success probability $p$. Upon success, the entangled link is immediately used to transmit qubits from the client to the server. While in the server, qubits (numbered grey circles) undergo decoherence (brown clock). Qubits are discarded from the server after they have existed for $w$ time steps. \label{fig:bqc_scenario}}
Our model of the quantum network architecture on which the BQC protocol is carried out is summarised in the bullet points below. A depiction is in Figure \ref{fig:bqc_scenario}. 

\begin{itemize}
    \item The server is honest, meaning that it carries out all tasks specified by the client. The BQC protocol protects against malicious server activity, as well as being robust to noise. Here, we solely aim to quantify the effect of noise on the protocol.
    \item Entanglement generation between client and server is performed with sequential attempts. Each attempt succeeds with probability $p$.
    \item Upon entanglement success, a qubit transmission procedure takes place. We assume that each qubit comes into existence in the server memory at the end of the corresponding time step.
    \item Immediately after transmission, each qubit is established with fidelity $F_{\text{est}}(p)$, where $F_{\text{est}} :[0,1] \rightarrow [0,1]$ is a decreasing function. In this way, we include a trade-off between rate and fidelity that is inherent to the entanglement generation process occurring between client and server. In the following, we choose this to be $F_{\text{est}}(p)=1- \lambda p$. Motivation for this choice of $F_{\text{est}}$ is given in Appendix \ref{sec:trade_off_appendix}.
    \item While they are stored in the server, qubits are subject to depolarising noise with a memory lifetime of $T$ time steps. For a $d$-dimensional density matrix $\rho \in \mathcal{D}(\mathcal{H}_d)$, this has action
\begin{equation}
\rho \rightarrow e^{-\frac{t}{T}}\rho + \big(1 - e^{-\frac{t}{T}}\big) \frac{\mathbb{I}_d}{d},
    \label{eqn:depolarising_noise}
\end{equation}
where $\mathbb{I}_d$ is the $d$-dimensional identity matrix and $t$ is the number of time steps for which $\rho$ has existed at the server. For the case of a qubit, i.e. $d=2$, the fidelity then decays as 
\begin{equation}
    F_{\text{est}} \rightarrow \big(F_{\text{est}}-\frac{1}{2}\big)e^{-\frac{t}{T}} + \frac{1}{2}.
\end{equation}
    \item To reduce decoherence, the server discards a qubit once it has been in memory for $w$ time steps.
    \item All local operations and measurements by the client and server devices are perfect and instantaneous. In particular, once all qubits required for the round are present in the server, it immediately and perfectly applies the measurement flow that has been specified by the client. 
    \item Before each round, the client chooses an element of $V$ uniformly at random. The corresponding qubit is the first one sent. The client then cycles through the qubits from $V$ in some pre-defined order. With this added randomness, the resulting order of the qubit ages will appear completely random. We continue with this assumption because it simplifies the resulting calculation of $p_{\text{av}}$, by removing any dependence of the qubit ages on events that occurred beyond the last $w$ time steps. More details of protocol test rounds are given in Appendix \ref{sec:poe_deriv}.
\end{itemize}
In our model, then, the fidelity of a qubit in the server depends only on the amount of time it has been stored there, and the entanglement generation success probability $p$. Notice that our set-up consists of the sequential attempted establishment of qubits at the server, and the discarding of these qubits after they have existed for a pre-defined number of time steps. We then have a situation analogous to that considered in the first part of this work, where the qubits function as the corresponding quantum resources. The methods given in Section \ref{sec:solutions} can then be applied to study this situation: the time taken to complete a round is $\tau_{(w,s)}$ time steps, where $s = |V|$ is the number of qubits required to produce a graph state, and $\tau_{(w,s)}$ is the waiting time as defined in Section \ref{sec:preliminaries}. Furthermore, the qubit fidelities at the time when the server applies its entangling gates and measurements are determined by the ending pattern $X_{(w,s)}$ which finishes the process. More specifically, it is possible to calculate $p_{\text{av}}$ exactly using the ending pattern distribution. We briefly describe this now.

Suppose that during a particular test round, at the time the server will carry out its local operations and measurements, the fidelities of the server qubits are $\vec{F} = (F_1,F_2,...,F_{|V|})$. Then, given the model described in the previous section, it is possible to find a function that tells us the probability of error of a test round,  $P_G(\vec{F})$. This is a polynomial in the values $F_i$, and has a form dependent on the graph $G$ and the choice of $k$-colouring. The details of how to obtain this function are given in Appendix \ref{sec:poe_deriv}. An expression for the \textit{average} probability of error of a test round is then
\begin{equation}
    p_{\text{av}} = \sum_{\vec{F}}  \mathbb{P}(\vec{F}) P_G ( \vec{F} ),
    \label{eqn:av_poe_1}
\end{equation}
where $\mathbb{P}(\vec{F})$ is the probability of obtaining the particular fidelity vector $\vec{F}$. Note that in the model introduced in the previous section, the qubit fidelities $\vec{F}$ are determined by the amount of time for which the qubits have been stored in the server. Moreover, recall that the ages of the links are contained exactly in the ending pattern $X_{(w,s)}$. Writing this dependence as 
\begin{equation}
   \vec{F} = \vec{F}(X_{(w,s)}),
\label{eqn:fidelity_vector_dependence}
\end{equation}
we then rewrite (\ref{eqn:av_poe_1}) to obtain an expression for the average probability of test round failure, 
\begin{equation}
    p_{\text{av}} =  \sum_{x \in \Omega(w,s)}  \mathbb{P}(X_{(w,s)} = x) P_G\left( \vec{F}(x) \right).
    \label{eqn:av_poe}
\end{equation}
This is a quantity that we can now evaluate using the methods introduced in Section \ref{sec:solutions}. In this way, the tools from Section \ref{sec:solutions} allow for the direct connection between the feasibility of the BQC protocol, as determined by $p_{\text{av}}$, to its rate. Since the above formula for $p_{\text{av}}$ is dependent on the graph structure and $k$-colouring, some parameter regimes may be sufficient for some calculations but not others. For example, for more complicated graphs that require a larger $k$, the condition (\ref{eqn:avg_error_condition}) is more strict. Further, if one chooses a different graph or $k$-colouring for the calculation, the polynomial $P_G$ may differ.

\subsubsection{Numerical evaluation}
\label{sec:bqc_results}
We now aim to find optimal values of the architecture parameters $p$ and $w$ for one round of the protocol. By \textit{optimality}, we mean that the expected time taken to carry out a round is minimised, while ensuring that the protocol is still feasible. Note that this does not necessarily mean optimality for the full protocol, which is comprised of multiple rounds. To optimise over the full protocol, one would to do a further optimisation over more protocol parameters (for example, the ratio of computation and test rounds), which we not not consider in this work. 

There is a combination of trade-offs between rate and fidelity present in our scenario: firstly due to varying the success probability, and secondly due to varying the window size. An increase in $p$ increases the rate at which successful links are generated, but decreases the initial fidelity of qubits in the server by an amount determined by $F_{\text{est}}(p)$. We would therefore expect that a smaller value of $w$ is required to minimise decoherence at the server, to ensure that the condition (\ref{eqn:avg_error_condition}) is met.  This in turn increases the expected time taken to generate all necessary entangled links within the time window. More formally, given a fixed $p$, we may find the minimal expected time for one round with the following procedure.

\begin{enumerate}
    \item Find the maximum value of $w$ such that the protocol is still feasible for this value of $p$, 
    \begin{equation}
        w_{\text{max}}(p) \coloneqq \text{max} \left\{w: p_{\text{av}} < \frac{2\gamma-1}{k(2\gamma-2)} \right\}.
        \label{eqn:max_w_bqc}
    \end{equation}
    \item Compute $\mathbb{E}\left(\tau_{(w_{\text{max}}(p),s)}\right)\big|_{p}$. This is the minimum expected time for one round.
\end{enumerate}

As an example of this method put into practice, we consider the case where the client would like to perform a BQP calculation on a square graph, so that $|V|=4$. This requires $s=4$ entangled pairs to be produced within the time window. For simplicity we will consider deterministic computations, so that the requirement on the probability of error is $p_{\text{av}}<1/2k$. In this case, $k$ can be chosen to be 2 (see Appendix \ref{sec:poe_deriv} for an example of a 2-colouring of a square graph), and the sufficient condition becomes $p_{\text{av}}<1/4$. 

\Figure[t!](topskip=0pt, botskip=0pt, midskip=0pt)[width=85mm]{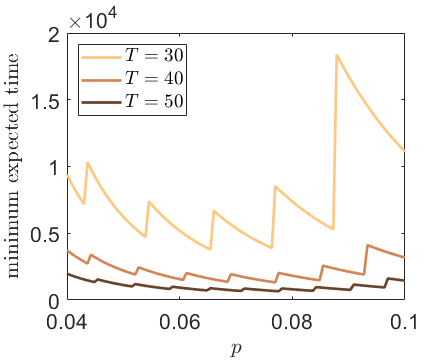}
{\textbf{Minimum expected time for one round of a BQC calculation vs. success probability of entanglement generation.} For a given value of $p$, we find the maximum window size (\ref{eqn:max_w_bqc}), and then use that to compute the minimum expected waiting time. Here, $p$ is discretised into 100 values of $p$ that are evenly spaced in the range $[0.04,0.1]$. We assume calculations on a square graph, which requires four entangled links to be produced within a time window. \label{fig:optimal_p_bqc}}

In Figure \ref{fig:optimal_p_bqc}, the minimum expected time to carry out a round is plotted against $p$ for three different values of the memory lifetime parameter $T$, and $F_0(p) = 1 - \lambda p$. We choose $\lambda = \frac{1}{2}$ in order to best display the behaviour of the solution, given our computational resources. In particular, the range of $p$ that we plot is chosen to clearly show the region of the optimal combination of the two trade-offs. For small $p$, the expected waiting time is high due to the small entanglement generation probability. For large $p$, it is high due to the small window size required due to the decrease in $F_\text{est}$. We therefore see a region in the middle of the plot where the average waiting time is minimal, or equivalently, the rate at which rounds can be carried out is maximal. For larger $T$, the decoherence of qubits in memory is reduced, and so it is possible to have a larger window size without disrupting the condition on $p_{\text{av}}$. We thus see that the expected time for one round decreases with $T$. Further, there are sharp peaks in the plots for each $T$, which are due to the discrete nature of $w$. This can be explained as follows: in the middle of two peaks, it is possible to increase the value of $p$ without disrupting the condition (\ref{eqn:avg_error_condition}). However, there will come a point where this condition is in fact an equality, which is when the window cut-off must be made smaller. This causes a discrete jump to a higher expected time.

\Figure[t!](topskip=0pt, botskip=0pt, midskip=0pt)[width=85mm]{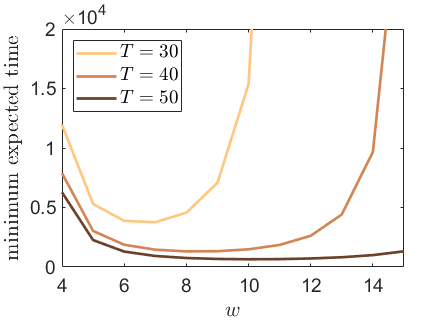}
{\textbf{Minimum expected time for one round of a BQC calculation vs. window size.} For each window size $4\leq w \leq 15$, we find the maximum success probability (\ref{eqn:max_p_bqc}), and then use this to compute the expected waiting time. We assume calculations on a square graph, which requires four entangled links to be produced within a time window.\label{fig:optimal_w_bqc}}

One can do something similar when varying the window size. Given a fixed $w$, we find the minimum expected time for one round with the following steps.
\begin{enumerate}
    \item Find the maximum value of $p$ such that the protocol is still feasible,
    \begin{equation}
        p_{\text{max}}(w) \coloneqq \text{sup} \left\{p:p_{\text{av}}<\frac{2\gamma-1}{k(2\gamma-2)} \right\}.
        \label{eqn:max_p_bqc}
    \end{equation}
    \item Compute $\mathbb{E}\left(\tau_{(w,s)}\right)\big|_{p_{\text{max}}(w)}$.
\end{enumerate}

In Figure \ref{fig:optimal_w_bqc}, the minimum expected time to carry out a round is plotted against the window size. This is again in the case of a square graph, for the same three values of the memory lifetime parameter $T$. We see a similar behaviour as when varying the success probability: a smaller $w$ induces a larger expected time. When $s$ is larger, qubits are subject to more decoherence, and in order to keep condition (\ref{eqn:avg_error_condition}) it is necessary to decrease the success probability. This is what induces a larger waiting time for larger $w$. We therefore again see an optimal region of $w$ for which the expected time to carry out one round of the protocol is minimised. 

Finally, we note that in practice, in order to optimise the full BQC protocol, one would need to consider how other aspects of the set-up, such as hardware, architecture and protocol, affect the performance. The simple scenario chosen in this work was to highlight the application of the results of Section \ref{sec:solutions}. We see from Figures \ref{fig:optimal_p_bqc} and \ref{fig:optimal_w_bqc} that for such values of $T$ and $s$, the methods from Section \ref{sec:solutions_approximations} enable one to make a careful choice of $(w,p)$ that can improve the rate of rounds of the protocol by two or three times, in comparison to other non-optimal choices of $(w,p)$ that are also sufficient for protocol feasibility.

\section{Further directions}
\label{sec:further_directions}
With the methods presented in this work, we focus on computing both the first and second moments of $\tau_{(w,s)}$, and the full distribution of $X_{(w,s)}$. We have seen that for $w$ finite and $s>2$, the formulae given here to compute $\mathbb{E}(\tau_{(w,s)})$ and the distribution of $X_{(w,s)}$ are in the form of linear systems that scale as $|\Omega(w,s)|+1$. If one would like to compute the full ending pattern distribution, then this seems to be a good scaling, since the outcome is comprised of $|\Omega(w,s)|$ probabilities. However, if one is for example only interested in $\mathbb{E}(\tau_{(w,s)})$ (e.g. for computing a protocol rate), then for certain regimes of $w$ and $p$ it may be useful to consider a continuous approximation, where the time between successful resource generation attempt is exponentially distributed. Such a case is often considered in the scan statistics literature (for example, see \cite{glaz2001scan}). However, the computation of the ending pattern in the continuous case is not immediately clear. 

We also note that a useful tool of approximation would be to further understand the asymptotic scaling hightlighted by Theorem \ref{thm:asymptotic_small_p}. More specifically, it would be interesting to know exactly how fast is the approach of (\ref{eqn:expectation_scaling_const}) and (\ref{eqn:ending_pattern_sim_uniform}), in terms of $s$ and $w$.

In the set-up of the problem, one could also consider a more realistic model of a quantum network architecture. For example, there may be parameter drift, when the success probability decreases over time due to increased noise. Further, in the more general case where the sequential attempts are not necessarily independent but Markovian, methods similar to those used in this paper may again be applied to the problem - see \cite{scan_statistics_methods_apps}, for example. 

\section*{Acknowledgement}
The authors thank Francisco Ferreira da Silva, Janice van Dam, and Álvaro G. Iñesta for critical feedback on the manuscript. BD acknowledges support from the KNAW Ammodo award (SW). BD, TB, and SW acknowledge support from the Quantum Internet Alliance (EU Horizon Europe grant agreement No. 101102140). GV was supported in part by the NWO ZK QSC Ada Lovelace Fellowship.

\appendices

\section{ (Identities for the case of two resource states)}
\label{sec:appendix_two_links}
\subsubsection{Evaluation of $\mathbb{E}(\tau_{(w,2)})$}
\begin{proof}[Evaluation of $\mathbb{E}(L)$]
Recalling from (\ref{eqn:ending_pattern_dist_s=2}) the distribution of $L$, we have 
\begin{align*}
    \mathbb{E}(L) &=\sum_{n=1}^{w-1} \frac{n(1-p)^{n-1}p}{1-(1-p)^{w-1}} \\ &= \frac{p}{1-(1-p)^{w-1}} \sum_{n=1}^{w-1} n(1-p)^{n-1} \\ 
    &= \frac{p}{1-(1-p)^{w-1}} \cdot - \dv{}{p} \sum_{n=1}^{w-1} (1-p)^{n} \\ 
    &= \frac{p}{1-(1-p)^{w-1}} \cdot - \dv{}{p} \frac{1-(1-p)^{w}}{1-p}
    \\ &= \frac{p}{1-(1-p)^{w-1}} \cdot \frac{1-(1-p)^w-wp(1-p)^{w-1}}{p^2} \\ 
     &= \frac{1-(1-p)^{w}-wp(1-p)^{w-1}}{p(1-(1-p)^{w-1})},
\end{align*}
where to evaluate the sum we have used the identity for a geometric series. 
\end{proof}

\begin{proof}[Proof that $\mathbb{E}\left( \sum_{j = 1}^M T_j \right)= \mathbb{E}(M)\mathbb{E}(T_1)$] This is used to evaluate the expectation $\mathbb{E}(\tau_{(w,2)})$. The random variables $M$ and $\{ T_j\}$ are independent, and since the $\{T_j\}$ are identically distributed,
\begin{align*}
    \mathbb{E}\left( \sum_{j = 1}^M T_j \right) & = \sum_{m=1}^{\infty} \mathbb{E}\left( \sum_{j = 1}^m T_j \right) \mathbb{P}(M=m) \\ & = \sum_{m=1}^{\infty} \mathbb{E}(T_1) \cdot m \mathbb{P}(M=m) \\ & = \mathbb{E}(T_1)\mathbb{E}(M).
\end{align*}

\end{proof}
    
\subsubsection{Evaluation of $\text{Var}(\tau_{(w,2)})$}
Recall $M$, $T_j$ and $L$, as given in Section \ref{sec:s=2}. These are independent, $M$ and $T$ have distributions $M\sim \text{Geom}(1-(1-p)^{w-1})$, $T_j \sim \text{Geom}(p)$, and $L$ has distrbiution as given in (\ref{eqn:ending_pattern_dist_s=2}). From (\ref{eqn:s=2_decomposition}), we have 
\begin{multline}
    \text{Var}(\tau_{(w,2)}) = \text{Var}\left( \sum_{j = 1}^M T_j  + (M-1)(w-1) \right) \\ + \text{Var}(L),
    \label{eqn:variance_decomp}
\end{multline}
since $L$ is independent of $M$ and $T_j$. Now, letting $\text{Cov}(X,Y) = \mathbb{E}(XY)-\mathbb{E}(X)\mathbb{E}(Y)$ be the covariance,
\begin{multline}
    \text{Var}\left( \sum_{j = 1}^M T_j  +  (M-1)(w-1) \right) \\ 
    = \text{Var}\left( \sum_{j = 1}^M T_j \right) + \text{Var} \left( (M-1)(w-1) \right) + \\ 2\cdot \text{Cov}\left(\sum_{j = 1}^M T_j,  (M-1)(w-1) \right),
    \label{eqn:s=2_variance_identity}
\end{multline}
where we have used the identity $\text{Var}(X+Y) = \text{Var}(X) + \text{Var}(Y) +2 \cdot\text{Cov}(X,Y) $.
We now evaluate (\ref{eqn:s=2_variance_identity}) term by term. Firstly,
\begin{align*}
     \mathbb{E} & \left(\! \left(  \sum_{j = 1}^M T_j \right)^2\right)  = \sum_m \mathbb{E}\! \left(\! \left( \sum_{j = 1}^m T_j \right)^2\right) \! \mathbb{P}(M\!=
    \!m) \\    
    &= \sum_m \mathbb{E}\left(\sum_{j=1}^m T_j^2+ \sum_{i\neq j} T_i T_j \right) \mathbb{P}(M\!=
    \!m) \\ 
    &= \sum_m \left( m\mathbb{E}(T_1^2) + m(m-1)\mathbb{E}(T_1)^2\right)\mathbb{P}(M\!=
    \!m) \\
    &= \mathbb{E}(M)\mathbb{E}(T_1^2)+\left(\mathbb{E}(M^2)-\mathbb{E}(M)\right)\mathbb{E}(T_1)^2.
\end{align*}

Subtracting $\mathbb{E}  \left( \sum_{j = 1}^M T_j \right)^2 = \mathbb{E}(M)^2\mathbb{E}(T_1)^2$ then yields
\begin{equation}
    \text{Var}\left( \sum_{j = 1}^M T_j \right) = \mathbb{E}(M)\text{Var}(T_1) + \text{Var}(M)\mathbb{E}(T_1)^2.
\end{equation}
Secondly, 
\begin{gather}
    \text{Var} \left( (M-1)(w\!-\!1) \right)\! = \!(w\!-\!1)^2\text{Var}(M).
\end{gather}
Thirdly,
\begin{align}
    \text{Cov} \Bigg(\sum_{j = 1}^M  T_j,  (M\!-\!1)(w\!-\!1) & \Bigg) \\  = (w\!-\!1) \text{Cov} \Bigg( & \sum_{j = 1}^M T_j,  M \Bigg) \\
     = (w-1) \Bigg( \sum_m m^2 \mathbb{E}(T_1)  \mathbb{P}(M=m ) - & \mathbb{E}(M)^2  \mathbb{E}(T_1) \Bigg) \\ 
    =(w-1)\text{Var}(M) \mathbb{E} (T_1) &.
\end{align}
It now remains to evaluate $\text{Var}(L)$. We firstly calculate 
\begin{gather}
    \mathbb{E}(L^2) = \sum_{n=1}^{w-1} \frac{n^2(1-p)^{n-1}p}{1-(1-p)^{w-1}} \\ = \mathbb{E}(L)+ \frac{p}{1-(1-p)^{w-1}} \sum_{n=1}^{w-1} n(n-1)(1-p)^{n-1}.
\end{gather}
Now,
\begin{gather*}
   \sum_{n=1}^{w-1} n(n-1)(1-p)^{n-1} = (1-p) \dv{^2}{p^2}\sum_{n=0}^{w-1} (1-p)^n \\ 
   = (1-p) \dv{^2}{p^2} \left(\frac{1-(1-p)^w}{p}\right),
\end{gather*}
so 
\begin{multline}
    \text{Var}(L) = \frac{p(1-p)}{1-(1-p)^{w-1}}  \dv{^2}{p^2} \left(\frac{1-(1-p)^w}{p}\right) \\ +\mathbb{E}(L) - \mathbb{E}(L)^2.
\end{multline}
We are now equipped to compute the full variance of $\tau_{(w,2)}$,
\begin{multline}
    \text{Var}(\tau_{(w,2)}) = \mathbb{E}(M)\text{Var}(T_1) + \text{Var}(M)\mathbb{E}(T_1)^2 \\ +2 (w-1)\text{Var}(M)\mathbb{E}(T_1) + (w-1)^2 \text{Var}(M) \\ + \text{Var}(L),
\end{multline}
where one may find a closed-form expression by inputting the standard identities for a geometric random variable, which are
\begin{align}
      & \mathbb{E}(T_1) =\frac{1}{p} \\
       & \text{Var}(T_1) = \frac{1-p}{p^2} \\
      & \mathbb{E}(M)=\frac{1}{1-(1-p)^{w-1}} \\
      & \text{Var}(M)=\frac{(1-p)^{w-1}}{(1-(1-p)^{w-1})^2}.
\end{align}

\section{(Ending pattern distribution and waiting time moments in the case of a finite window size)}
\subsubsection{The waiting time is well-defined}
\label{sec:waiting_time_well_defined}
    We show here that $\tau_x$ can be bounded above by a geometrically distributed random variable. Using the notation $p_1 = p$, $p_0 = 1-p$ for an ending pattern $x \in \Omega_l(s)$, this exact sequence will appear in any given $l$ consecutive trials $Z_{i-l+1},...,Z_{i}$ with probability $\gamma_x :=p_{x_1}...p_{x_l}$. Defining a new sequence of random variables $(Y_n)_{n=1}^{\infty}$,
    \begin{equation}
        Y_n = \begin{cases} 
         1 \text{ if } Z_i = x_i \text{ for all } (n-1)l < i \leq nl; \\
         0, \text{ otherwise.}
        \end{cases}
    \end{equation}
    Each $Y_n$ is then Bernoulli with parameter $\gamma_x$. It takes the value 1 if the $n$th segment of $l$ trials exactly matches with $x$. There is then an associated waiting time random variable $\tilde{\tau}_x$ that is geometric with parameter $\gamma_x$,
    \begin{equation}
        \tilde{\tau}_x:= \text{min} \{n: Y_n = 1 \}.
    \end{equation}
    Moreover, the waiting time to see $x$ satisfies $\tau_x \leq  \tilde{\tau}_x\cdot l$. Taking expectations yields 
    \begin{equation}
        \mathbb{E}(\tau_{(w,s)}) \leq \mathbb{E}(\tau_x) \leq \mathbb{E}(\tilde{\tau}_x \cdot l) = \frac{l}{\gamma_x} < \infty,
\label{eqn:finite_expectation}
    \end{equation}
which completes our proof. We note that the same method can be used to show that all moments of $\tau_{(w,s)}$ are finite.

\subsubsection{The expected waiting time of a simple pattern}

\label{sec:derivation_finite_window_size}
Using the theory of martingales and a helpful gambling analogy to aid understanding,  we now derive a way to numerically compute the ending pattern distribution $\{ \mathbb{P}(x): x \in \Omega(w,s) \}$, and the first and second moments of the waiting time $\tau_{(w,s)}$, in the case of a finite window size. The result of this is Theorem \ref{thm:exact_solutions} in the main text. The method was introduced in \cite{martingale_approach}, where they consider the more abstract case of a general sequence of discrete i.i.d random variables, and a general set of ending patterns. Here, due to its relevance to the subject of the main text, we continue with the case of i.i.d. Bernoulli trials.

\label{sec:appendix_simple_pattern}
It will be useful to first of all consider the case where we wait for an instance of a single pattern $x = (x_1,...,x_l)\in \{0,1 \}^l$, instead of waiting for any instance of the set $\Omega(w,s)$. The former case is referred to as a \textit{simple pattern} and the latter as a \textit{compound pattern}. In this section we will find an exact expression for $\mathbb{E}(\tau_x)$. Here, $\tau_x$ refers to the waiting time until seeing the pattern $x$, and is defined in (\ref{eqn:wt_simple_def}).

To provide intuition, we introduce the following scenario of gamblers in a casino. Suppose that just before the first trial is realised, a gambler, hereinafter referred to as Gambler 1, bets \EUR $1$ on the outcome $\{Z_1 = x_1 \}$. We also suppose that the odds are fair, so that if this is the case then she wins \EUR $\frac{1}{p_{x_1}}$ \footnote{i.e. the expected net gain of the gambler is zero. Calling this G, we can verify explicitly by writing $\mathbb{E}(G) = (1/p_{\lambda_1} -1)\cdot p_{\lambda_1}+ (-1)\cdot (1-p_{\lambda_1}) = 0.$}. Moreover, if she wins, then she straight away bets all of these winnings on the outcome $\{Z_2 = x_2 \}$.  If not, the casino keeps her \EUR $1$ and she doesn't place any more bets. For a general $n$, Gambler 1 then proceeds at the $n$th trial in a similar way: if she has yet to lose, she bets all of her winnings on the outcome $\{Z_n = x_n \}$, and if not, she doesn't place any bet. Furthermore, at every trial we introduce a new gambler who behaves in exactly the same way, so that Gambler $2$ bets \EUR $1$ on the outcome $\{Z_2 = x_1 \} $, and continues betting all of her winnings on the subsequent rounds being equal to the next entry of $x$, up until she loses a round. Gambler $j$ bets \EUR $1$ on the outcome $\{Z_j = x_1 \}$ and continues with exactly the same strategy. The game stops when the sequence $x$ first appears, which by definition is at the $\tau_x$th trial.

Our aim now is to write down an expression for the combined net gain of the gamblers after the $n$th trial, for a general $n$. In order to do this concisely, we recall the definition (\ref{eqn:delta_definition}) of the quantities $\delta_{(a,b)}$. Given a realisation $C_n := (c_1,...,c_n)$ of the first $n$ trials, the winnings of Gambler $j$ after the $n$th round can then be written as 
\begin{equation}
    W^{(j)}(C_n) = \begin{cases} \delta_{(x_1,c_j)} \delta_{(x_2,c_{j+1})} ... \delta_{(x_{n-j+1},c_n)} \\  \text{ for } n-l+1 \leq j \leq n; \\
    0 \text{, otherwise.}
    \end{cases}
    \label{eqn:winnings_gambler_j_simple}
\end{equation}
Here, we see that the functions $\delta_{(a,b)}$ allow us to elegantly write down a Gambler's winnings. With this in mind, the \textit{combined} winnings of the gamblers after the $n$'th trial is
\begin{equation}
    W(C_n) : = \sum_{j = 1}^n W^{(j)}(C_n) = \sum_{j = n-l+1}^n W_n^{(j)} \equiv x * C_n,
    \label{eqn:total_winnings_simple}
\end{equation}
where we have introduced the function $*$ that was defined in (\ref{eqn:star_def}), and that maps two binary strings to a scalar value. From (\ref{eqn:winnings_gambler_j_simple}), we see that the \textit{net gain} of the $j$th gambler after the $n$th time-step is simply 
\begin{equation}
    G^{(j)}(C_n):= W^{(j)}(C_n) - 1, \label{eqn:net_gain_gambler_j_simple}
\end{equation}
and similarly, the \textit{total net gain} of the gamblers after the $n$th trial is
\begin{equation}
    G(C_n) := \sum_{j = 1}^n G^{(j)}(C_n) \\  = x*C_n - n.
    \label{eqn:total_net_gain_simple}
\end{equation}
We can now define a sequence of random variables $(G_n)_{n\geq0}$,
\begin{equation}
    G_n:= G(C_n)
\end{equation}
which take the value of the total net gain of the gamblers after each round. In particular, after the game ends, the total net gain is
\begin{equation}
    G_{\tau_x} = x*x - \tau_x.
    \label{eqn:net_gain_final_simple}
\end{equation}
Note that $x*x$ is a quantity that is only dependent on the pattern $x$. Since the game is defined to be fair at every round, the expected total net gain when the game finishes would intuitively be equal to zero, i.e. 
\begin{equation}
    \mathbb{E}(G_{\tau_x}) = 0,
\label{eqn:exp_is_0_simple_pattern}
\end{equation}
A neat expression for the expected waiting time to see the sequence $B$ follows by making use of the linearity of expectation,
\begin{equation}
    \mathbb{E}(\tau_x) = x*x.
    \label{eqn:exp_wt_simple}
\end{equation}
To prove (\ref{eqn:exp_is_0_simple_pattern}), we make use of the fact that $(G_n)_{n\geq 0}$ is a \textit{martingale}, for which the following properties must hold:
\begin{enumerate}[label=(\roman*)]
\item $\mathbb{E}(|G_n|)<\infty.$
\item $\mathbb{E}(G_{n+1}|G_n,...,G_1) =  G_n$
\end{enumerate}
To show (i), we use the definition (\ref{eqn:net_gain_final_simple}), and see that
\begin{equation}
        \mathbb{E}(|G_{n}|) \leq x*C_n + \mathbb{E}(\tau_x) < \infty,
\end{equation}
since the waiting time $\tau_x$ is well-defined, and $x*C_n$ is bounded. To show condition (ii), we use the fact that the game is fair at each round. Suppose that we have the maximum amount of information about what has happened in the first $n$ trials, i.e. we know that they have taken the values $(c_1,...,c_n)$. Then, the conditional expectation of $G_{n+1}$ satisfies
\begin{gather*}
    \mathbb{E}(G_{n+1}|Z_1=c_1,...,Z_n=c_n) \\
    = \sum_{c_{n+1} \in \{0,1\}} G\left(C_n,c_{n+1}\right) \mathbb{P}(Z_{n+1} = c_{n+1}) \\ = \sum_{j=1}^{n+1} \sum_{c_{n+1} \in \{0,1\}} G^{(j)}\left(C_n,c_{n+1}\right) p_{c_{n+1}} \\ = \sum_{j=1}^{n+1} G^{(j)}(C_n) = G_n,
\end{gather*}
where to go to the final line, we have made use of the definition of $G^{(j)}$. Since the realisations of $(Z_1,...,Z_n)$ completely determine the values of $G_1,...,G_n$, this also shows (ii).

We now know that $(G_n)_{n\geq 0}$ is a martingale. However, this is not quite enough to show (\ref{eqn:exp_is_0_simple_pattern}), which is what is required to obtain the final simple form for $\mathbb{E}(\tau_x)$. In particular, some extra machinery is needed, in the form of Doob's \textit{optional stopping theorem}, a proof of which can be found in \cite{probability_with_martingales}. A version of this is stated below.
\begin{theorem}[Optional stopping]
    Let $G_n$ be a martingale and $\tau$ a stopping time. Suppose that there exists a constant $K$ such that $|G_n - G_{n-1}|<K$ for all $n$. Suppose also that $\tau$ is a.s. finite. Then $\mathbb{E}(G_{\tau}) = \mathbb{E}(G_1)$.\label{thm:doob}
\end{theorem} 
All that remains to be done is to show that the martingale defined in (\ref{eqn:net_gain_final_simple}) satisfies the required properties to satisfy Theorem \ref{thm:doob}. Firstly, we have 
\begin{gather}
    |G_{n}-G_{n-1}| < x*C_n + x*C_{n-1} + 1\leq K,
\end{gather}
where 
\begin{equation}
    K = 2\cdot \text{max}_{C \in \{0,1 \}^l} \{x*C\} +1.
\end{equation}
Secondly, we see that since $\tau_x$ is bounded above by a geometric random variable, it is a.s. finite. This gives us (\ref{eqn:exp_is_0_simple_pattern}).

\subsubsection{Starting from another pattern}
We now adapt the results above in order to find the expected time to see $x$, given that we start already with some pattern $y$.
We extend the gambling analogy in order to illustrate this concept, and suppose that we want to calculate the expected time until seeing $y$ only \textit{after} some number of rounds, $m$, say, have been realised. In particular, after the $m$th round we know the first $m$ realisations of the i.i.d Bernoulli sequence, and we call these $y = (y_1,...,y_m)$. At this point, the net gain of the gamblers is thus $G_m = x*y - m$. We will evaluate the net gain of the gamblers \textit{compared} to this point after each of the $n$ trials, which for $n \geq m$ we denote by $\tilde{G}_n$. This is simply given by 
\begin{gather}
    \tilde{G}_n = G_n - G_m = (x*C_n - n) - (x*y - m) \\ = x*C_n - x*y - (n-m),
\end{gather}
where $C_n$ is no longer completely general as its first $m$ entries must correspond to $y$. Using the same reasoning as before, one can show that $(\tilde{G}_n)_{n\geq 0}$ is a martingale. Then, defining $\tau_{xy}$ as the waiting time to see $x$ \textit{given} that we have already seen pattern $y$, it is again possible to use Theorem \ref{thm:doob} to show that
\begin{equation*}
    0 = \mathbb{E}(\tilde{G}_{\tau_{xy}}) = \mathbb{E}(x*x - x*y - \tau_{yx}),
\end{equation*}
and so by the linearity of expectations,
\begin{equation}
    \mathbb{E}(\tau_{xy}) = x*x - x*y.
    \label{eqn:exp_wt_start_another_pattern}
\end{equation}
We may now use the results derived above to derive a formula for $\mathbb{E}(\tau_{(w,s)})$ and the distribution of $X_{(w,s)}$. Given $x \in \Omega(w,s)$, we write
\begin{gather*}
    \mathbb{E}(\tau_x) = \mathbb{E}(\tau_{(w,s)}) + \mathbb{E}(\tau_x - \tau_{(w,s)})
   \\ = \mathbb{E}(\tau_{(w,s)}) \! +\! \sum_{y \in \Omega(w,s)}\! \mathbb{P}(X_{(w,s)} = y) \mathbb{E}(\tau_x \! - \! \tau_{(w,s)}|X_{(w,s)} = y) \\ = \mathbb{E}(\tau_{(w,s)}) + \sum_{y \in \Omega(w,s)} \mathbb{P}(X_{(w,s)} = y) (x*x - x*y),
\end{gather*}
Where we have noticed that  $\mathbb{E}(\tau_x - \tau_{(w,s)}|X_{(w,s)} = y)  = \mathbb{E}(\tau_{xy})$.
Applying Theorem \ref{thm:doob} and enforcing the condition that the ending pattern probabilities must sum to one then yields the formula (\ref{eqn:linear_system}).

\subsubsection{Formula for the second moment of the waiting time}
\label{sec:second_moment}
An extension to the gambling analogy given above can be used to derive the formula for the second moment of the waiting time, for which we refer to \cite{scan_statistics_methods_apps}. Here, we will only state the formula. We first of all define a new operation $\dagger$ that maps two elements $x,y \in \Omega(w,s)$ to a real number. If $x = (x_1,...,x_k)$ and $y = (y_1,...,y_m)$, 
\begin{equation}
    x \dagger y := \sum_{j = 1}^{\text{min}(k,m)}  (1-j)\prod_{i = 1}^{j} \delta_{(x_i,y_{m-j+i})} .
\end{equation}
Letting $\Omega \equiv \Omega(w,s)$, the second moment of $\tau_{(w,s)}$ can be found through solving the following systems
\begin{theorem}
Let $\{u_j \}_{1\leq j \leq |\Omega|}$ and $\{v_j \}_{1\leq j \leq |\Omega|}$ solve the linear systems
\begin{gather}
    \sum_{j=1}^{|\Omega|}W_{ij} u_j = 1, \text{ for } 1 \leq i \leq |\Omega|, \\
    \sum_{j=1}^{|\Omega|} (N_{ij}u_j + W_{ij}v_j) = 1, \text{ for } 1 \leq i \leq |\Omega|
\end{gather}
with $W_{ij}:= x^{(i)} * x^{(j)}$ and $N_{ij} := x^{(i)} \dagger x^{(j)}$. Then,
\begin{equation}
\mathbb{E}\left(\!\tau_{(w,s)}^2\!\right)\! = \!\frac{1+ \left(1\!-\!\sum_j v_j \! - \!\sum_j u_j/2 \right)\cdot \mathbb{E}(\tau_{(w,s)})}{\sum_j u_j/2}.
\end{equation}
\end{theorem}
Code that makes use of this formula to compute $\mathbb{E}\left(\tau_{(w,s)}^2\right)$ is provided in \cite{github_repository}.
\section{(Approximations)}
\subsubsection{Infinite window size approximation}
\label{sec:appendix_sum_evaluation} 
\begin{proof}[Proof of Theorem \ref{thm:infinite_w_bound}]
Letting $\epsilon \equiv \epsilon(w,s,p)=\mathbb{P}(\tau_{(w,s)}>w)$, the expectation of $\tau_{(w,s)}$ can be rewritten as
\begin{multline}
\label{eqn:step_1}
    \mathbb{E}(\tau_{(w,s)}) = (1-\epsilon)\mathbb{E}\left(\tau_{(w,s)}|\tau_{(w,s)}\leq w\right) + \\ \epsilon \mathbb{E}\left(\tau_{(w,s)}|\tau_{(w,s)}> w\right).
\end{multline}
Now, note that for $n\leq w$ 
\begin{equation*}
    \mathbb{P}(\tau_{(w,s)}=n) = \mathbb{P}(\tau_{(\infty,s)}=n),
\end{equation*}
 i.e. for this range of $n$ the distributions of $\tau_{(w,s)}$ and $\tau_{(\infty,s)}$ exactly match. We can thus rewrite (\ref{eqn:step_1}) as
    \begin{multline*}
      (1-\epsilon)\mathbb{E}\left(\tau_{(\infty,s)}|\tau_{(\infty,s)}\leq w\right) + \epsilon \mathbb{E}\left(\tau_{(w,s)}|\tau_{(w,s)}> w\right) \\ = \mathbb{E}(\tau_{(\infty,s)}) - \epsilon \mathbb{E}\left(\tau_{(\infty,s)}|\tau_{(\infty,s)}>w\right) + \\\epsilon \mathbb{E}\left(\tau_{(w,s)}|\tau_{(w,s)}> w\right),
\end{multline*}
where to obtain the last equality we have expanded $\mathbb{E}(\tau_{(\infty,s)})$ in the same way as (\ref{eqn:step_1}). Now, if one considers starting the whole process again after the first $w$ time steps, we see that $\mathbb{E}\left( \tau_{(w,s)}|\tau_{(w,s)}> w \right) \leq w+\mathbb{E}(\tau_{(w,s)})$. Combining this with the fact that $\mathbb{E}(\tau_{(\infty,s)}|\tau_{(\infty,s)}>w)>w$, we find that 
\begin{equation}
    \mathbb{E}(\tau_{(w,s)})-\mathbb{E}(\tau_{(\infty,s)}) \leq \epsilon \cdot \left(w+ \mathbb{E}(\tau_{(w,s)}) - w\right),
\end{equation}
from which (\ref{eqn:difference_bound}) follows.

We further bound the distance between the ending pattern distributions. Making use of 
\begin{multline*}
    \mathbb{P}(X_{(w,s)}=x) = \mathbb{P}(X_{(w,s)}=x|\tau_{(w,s)}\leq w)(1-\epsilon) \\  +\mathbb{P}(X_{(w,s)}=x|\tau_{(w,s)}> w)\epsilon,
\end{multline*}
we have 
\begin{multline*}
   \mathbb{P}(X_{(w,s)}\!=\!x)-\mathbb{P}(X_{(\infty,s)}\!=\!x) \\ = \big(\mathbb{P}(X_{(w,s)}\!=\!x|\tau_{(w,s)}\!>\!w)
     -\mathbb{P}(X_{(\infty,s)}\!=\!x|\tau_{(w,s)}\!>\!w)\big)\epsilon,
\end{multline*}
and so, letting $\Omega \equiv \Omega(\infty,s)$
\begin{multline*}
    \sum_{x\in \Omega}\! |\mathbb{P}(X_{(w,s)}\!=\!x)-\mathbb{P}(X_{(\infty,s)}\!=\!x)| <
     \\  \sum_{x\in \Omega}\! \bigg(\mathbb{P} (\!X_{(w,s)}\!=\!x|\tau_{(w,s)}\!>\!w)+\mathbb{P}(X_{(\infty,s)}\!=\!x|\tau_{(w,s)}\!>\!w)\!\bigg)\epsilon \\ = 2\epsilon(w,s,p).
\end{multline*}
\end{proof}

\begin{proof}[Proof of Lemma \ref{lem:epsilon_sum}]
Here, we show the identity (\ref{eqn:epsilon_identity}) for $\epsilon(w,s,p)$. Letting $q=1-p$, we have 
\begin{align*}
    \epsilon(w,s,p) &= \sum_{n = w+1}^{\infty} {n-1 \choose s-1}q^{n - s}p^{s} \\ &= \frac{p^s}{(s-1)!} \sum_{n = w+1}^{\infty} \frac{(n-1)!}{(n-s)!} q^{n-s} \\ &= \frac{p^s}{(s-1)!} \sum_{n = w+1}^{\infty} \dv{^{s-1}}{q^{s-1}} \left(q^{n-1}\right) \\ &= \frac{p^s}{(s-1)!} \dv{^{s-1}}{q^{s-1}} \left( \sum_{n = w+1}^{\infty} q^{n-1} \right) \\ &= \frac{p^s}{(s-1)!} \dv{^{s-1}}{q^{s-1}} \left( \frac{q^w}{1-q} \right) \\ = \frac{p^s}{(s-1)!} & \sum_{i=1}^{s-1} {s-1 \choose i} \dv{^i}{q^i} \left(q^w \right)  \dv{^{s-1-i}}{q^{s-1-k}}  \left((1-q)^{-1} \right) \\ = \frac{p^s}{(s-1)!} & \sum_{i=1}^{s-1} \frac{(s-1)!}{i! (s-i-1)!} \frac{w!}{(w-i)!} q^{w-i} \frac{(s-i-1)!}{(1-q)^{s-i}} \\ &= \sum_{i=1}^{s-1} {w\choose i} q^{w-i} p^i.
\end{align*}
\end{proof}
\label{sec:decreasing_fn_appendix}
\subsubsection{Asymptotic behaviour of the expected waiting time and ending pattern}
\label{sec:appendix_asymptotics}

\begin{proof}[Proof of Theorem \ref{thm:asymptotic_small_p}] 
     For conciseness, here we take $\Omega \equiv \Omega(w,s)$. A formula for the inverse of $A$ is given in terms of its adjugate matrix $\adj A$ \cite{horn_johnson_1985},
\begin{equation}
    [ \adj A ]_{ij} := (-1)^{(i+j)} \det M_{ji}.
\end{equation}
where $M_{ij}$ is the $|\Omega|\times |\Omega| $ matrix obtained by removing row $i$ and column $j$ from $A$. Since $A$ is invertible, the inverse is 
    \begin{equation}
        A^{-1} = \frac{\adj A}{\det A },
        \label{eqn:inverse_matrix_formula}
    \end{equation}
    Now consider the system (\ref{eqn:linear_system}). If we consider solving for $\vec{v}$ by multiplying through by $A^{-1}$, we see that its first element is
    \begin{equation}
        \mathbb{E}(\tau_{(w,s)}) = \frac{\det B }{\det A },
        \label{eqn:expectation_as_det}
    \end{equation}
    where $B$ is the $|\Omega| \times |\Omega|$ matrix obtained by removing the first row and column from $A$, so that $B_{ij} = x^i * x^j$. Since all the entries of $A$ are polynomials in $1/p$ and $1/q$, so are $\det B$ and $\det A$. 
    
    To proceed with showing (\ref{eqn:expectation_scaling_const}), we characterise the scaling of $\det B $ and $\det A $ for small $p$. Since $q=1-p$ is close to $1$ for small $p$, it suffices to only consider the powers of $1/p$ for the analysis of the asymptotic scaling as $p\rightarrow 0$ (recalling the definition of the star product (\ref{eqn:star_def})). We firstly consider $\det B$. With the observation that the higher-order terms in $1/p$ are given by the star products on the diagonal, and moreover that these each have leading order term given by $1/p^s$. The form of $\det B $, then, is a polynomial of maximum degree $1/p^{s|\Omega|}$. In fact, this is exactly the degree. One can compute this contribution by considering the matrix $\tilde{B}$ of highest powers: letting $r \equiv 1/p^s$, we have $\det B \sim \det \tilde{B}$, where
    \begin{equation}
        \tilde{B} = \left( 
        \begin{array}{cccc}
            r & 0 & ... & 0 \\
            0 & r & ... & 0 \\
            \vdots & & \ddots & 0 \\
            0 & 0 & ... & r
        \end{array}
        \right),
    \end{equation}
and hence, $\det B \sim r^{|\Omega|} = 1/p^{s|\Omega|}$. We then do the same with $A$. In this case, $\det A \sim \det \tilde{A}$, where 
\begin{equation}
    \tilde{A} = \left( 
        \begin{array}{ccccc}
         0 & 1 & 1 & 1 & 1 \\
        -1 & r & 0 & ... & 0 \\
        -1 & 0 & r & ... & 0 \\
        \vdots & \vdots & & \ddots & 0 \\
        -1 & 0 & 0 & ... & r
        \end{array}
        \right),
\end{equation}
where the existence of a $0$ in the top left-hand corner now disrupts the evaluation of $\det \tilde{A} $ by multiplying along the diagonal, as we did above. Our next step is to evaluate $\det \tilde{A} $ by expanding along the top row, 
\begin{equation}
    \det \tilde{A} = \sum_{k=1}^{|\Omega|} (-1)^k \det \tilde{A}_k,
    \label{eqn:det_X_expansion}
\end{equation}
where $\tilde{A}_k$ is the $|\Omega|\times |\Omega|$ matrix formed by removing the first row and the $k$th column from $\tilde{A}$, 
\begin{equation}
    \arraycolsep=4pt
     \def\arraystretch{1.2}
    \tilde{A}_k = \left(
    \begin{array}{ccccccc}
        -1  & r & 0 &   &  \hdots  &  & 0 \\
    \vdots  & 0 &  \ddots & &  & &  \vdots \\
            &   &   &  r &  0  & &   \\
            &   &   &  0 &  0  & &  \\
            &   &   &  0 & r &        &   \\
    \vdots  &   &   &    &   & \ddots & 0 \\
        -1  & 0 &\hdots&    & \hdots & 0 & r 
    \end{array}        
        \right).
\end{equation}
In $\tilde{A}_k$, the $r$'s are placed above the diagonal in rows $1,...,k-1$, and on the diagonal in rows $k+1,...,|\Omega|$. The determinant of $\tilde{A}_1$ may be calculated by simply multiplying the diagonal elements, to obtain 
\begin{equation}
    \det \tilde{A}_1 = -r^{|\Omega|-1}.
\end{equation}
We then notice that any $\tilde{A}_k$ can be transformed into $\tilde{A}_1$ by moving the $k$th row to the top row. This can be achieved by performing $k-1$ row operations, if it is moved by successively exchanging with the row above it $k-1$ times. Then, since each row operation incurs a factor of $(-1)^k$,
\begin{equation}
    \det \tilde{A}_k = (-1)^{k-1}\det \tilde{A}_1 = (-1)^k r^{|\Omega|-1}.
\end{equation}
With (\ref{eqn:det_X_expansion}), we then see that 
\begin{equation}
    \det \tilde{A} = |\Omega| r^{|\Omega|-1},
    \label{eqn:det_x_soln}
\end{equation}
and so $\det A \sim |\Omega| /p^{s(|\Omega| - 1)}$.
Substituting into (\ref{eqn:expectation_as_det}), we find
\begin{equation}
    \mathbb{E}(\tau_{(w,s)}) \sim \frac{p^{s(|\Omega|-1)}}{|\Omega|p^{s|\Omega|}} = \frac{1}{|\Omega|p^s}.
\end{equation}
To show (\ref{eqn:ending_pattern_sim_uniform}), we employ a similar method. We have from (\ref{eqn:inverse_matrix_formula}) that
\begin{gather}
    \mathbb{P}(X_{(w,s)}=x^{(k)}) = (-1)^{1+k}\frac{\det M_{k0}}{\det A} \\ \sim (-1)^{1+k}\frac{\det C_k}{\det X},
\end{gather}
where $C_k$ is obtained by removing the first column and $k$th row from $A$,
\begin{equation}
    C_k = \left(
    \arraycolsep=4pt
     \def\arraystretch{0.6}
    \begin{array}{ccccccc}
                     1 &\hdots &       &       &       & \hdots& 1 \\
                     r & 0     &\hdots &\pddots&       &\hdots & 0 \\    
                     0 & \ddots&       &       &       &\pddots&\vdots\\ 
                 \vdots&       & r     &  0    &  0    &       &\pddots\\
                \pddots&\pddots&   0   &  0    & r     &       &\vdots\\
                 \vdots&       &       &       &       &\ddots & 0 \\
                     0 &\hdots &\pddots&       &\hdots & 0     & r 
    \end{array}        
        \right),
\end{equation}
where in $C_k$ the $x$-entries are placed below the diagonal in columns $1,...,k-1$, and on the diagonal in columns $k+1,...,|\Omega|$. We note that $C_k$ is the transpose of $B_k$, with the first column multiplied by a scaling factor of $-1$. Therefore, $\det C_k = -\det B_k = (-1)^{k+1}r^{|\Omega|-1}$, and making use again of (\ref{eqn:det_x_soln}), 
\begin{equation}
    \mathbb{P}(X_{(w,s)}=x^{(k)}) \sim \frac{r^{|\Omega|-1}}{|\Omega|r^{|\Omega|-1}},
\end{equation}
from which the result follows.
\end{proof}

\section{(Trade-off function due to entanglement generation scheme)}
\label{sec:trade_off_appendix}
In this appendix, we motivate the linear trade-off function $F_{\text{est}}=1-\lambda p$ used in the BQC analysis, which describes the fidelity of qubits in the server immediately after transmission.

Let $\sigma$ denote the noisy two-qubit entangled state that is produced between the client and server when there is a successful attempt. When there is a success, the client and server perform some qubit transmission procedure $\Lambda_{\sigma}$, which could for example be teleportation, or a remote state preparation protocol. We assume that this protocol establishes all qubits in the server with the same fidelity $F_{\text{est}}$. For example, this is the case if the noisy entangled state is depolarised
\begin{equation}
    \sigma = \frac{4F_0-1}{3} |\Phi^+ \rangle \langle \Phi^+ | + \frac{1-F_0}{3} \mathbb{I}_4,
    \label{eqn:depolarised_pair}
\end{equation}
and the standard teleportation protocol from \cite{original_teleportation} is applied. Here, $F_0=\bra{\Phi^+} \sigma \ket{\Phi^+}$ is the fidelity of $\sigma$ to the target state. This involves performing a full measurement in the Bell basis $\{ |\Phi_{ij} \rangle\}$ and applying the corresponding Pauli corrections. If $\ket{\psi}$ is the qubit state to be teleported, its action is given by $\Lambda^{\text{st}}$, 
\begin{equation}
    \Lambda^{\text{st}}_{\sigma}(|\psi \rangle \langle \psi |)\! := \!\sum_{i,j} X^i Z^j \langle \Phi_{ij}| (|\psi \rangle \langle \psi | \otimes \sigma)| \Phi_{ij} \rangle Z^j X^i
    \label{eqn:teleportation_action}
\end{equation}
where the Bell measurement acts on the registers containing the qubit state $|\psi \rangle$ and the first qubit of $\sigma$. Suppose that the entangled state and qubit transmission procedure are given by (\ref{eqn:depolarised_pair}) and (\ref{eqn:teleportation_action}). Then after transmitting any qubit $\ket{\psi}$, the resulting fidelity is \cite{horodeckis_teleportation}
\begin{equation}
  F_{\text{est}} = \frac{2F_0+1}{3}.
  \label{eqn:relationship_fidelities}
\end{equation}
Now, one can incorporate a general rate-fidelity trade-off inherent to the entanglement generation protocol by specifying that $F_{\text{est}}$ is a decreasing function of $p$. In particular, we draw here on an example from the single-photon scheme for entanglement generation.  When implementing a single-photon scheme \cite{cabrillo1999creation}, the fidelity of generated states is 
\begin{equation}
    F_0(p_{\text{suc}}) = 1-\frac{p_{\text{suc}}}{2p_{\text{det}}},
    \label{eqn:fidelity_NV_single}
\end{equation}
where $p_{\text{suc}}$ is the success probability of a physical entanglement attempt, and $p_{\text{det}}$ is the probability of detecting an emitted photon. In the case of a very small $p_{\text{suc}}$, one might want to perform entanglement attempts in \textit{batches} in order to minimise overhead due to communication with higher layers of the software stack (which must be notified when there is, or is not, a success). This scheme has been implemented with NV centres in diamond, where typically $p_{\text{suc}} \ll 1$ \cite{single_click}. If this is the case,  choosing one time step to correspond to a batch of $M \ll 1/(p_{\text{suc}})^2$ attempts, the probability of producing at least one entangled link in a time step is
\begin{equation}
    p = 1 - (1-p_{\text{suc}})^M \approx M p_{\text{suc}}.
\end{equation}
Substituting this into (\ref{eqn:relationship_fidelities}) and (\ref{eqn:fidelity_NV_single}), we obtain a trade-off function of
\begin{equation}
    F_{\text{est}}(p) = \frac{2 \left(1-\frac{p}{2M p_{\text{det}}}\right)+1}{3} = 1 - \lambda p,
    \label{eqn:rate_fid_tradeoff_linear}
\end{equation}
where $\lambda := 1/(3Mp_{\text{det}})$. Since $M$ is a freely adjustable parameter, then so is $\lambda$. The simple relationship (\ref{eqn:rate_fid_tradeoff_linear}) is also a general first-order behaviour for a decreasing function in $p$ such that $F_{\text{est}}(0) = 1$, which justifies the choice as potentially applicable to other hardware and entanglement generation protocols.

\section{(Computing the error probability of a BQC test round)}
\label{sec:poe_deriv}
\subsubsection{Test rounds}
\label{sec:appendix_test_rounds}
As mentioned previously, the protocol involves interweaving test rounds at random with computation rounds. It is the test rounds that provide verifiability of the protocol, because they allow the client to check for deviations from the ideal measurement outcomes. Recall that the goal of the client in the BQC protocol is to perform a BQP computation, which is defined in the measurement-based formalism with respect to a graph $G = (V,E)$. In one round of the protocol, the client transmits $|V|$ qubits to the server, which (if it is honest) creates a graph state by applying $CZ$-gates to pairs of qubits as given in the set of edges $E$. Before carrying out the protocol, the client chooses some \textit{$k$-colouring} $\{V_j:j = 1,...,k \}$, which is a partition of the set of vertices $V$ into different subsets, known as \textit{colours}, such that there is no edge between two vertices of the same colour. This $k$ then corresponds to the $k$ in the feasibility condition (\ref{eqn:avg_error_condition}). 

Before each test round, the client chooses a colour $V_j$ uniformly at random to be the \textit{trap colour}. A qubit corresponding to vertices from this set is then referred to as a \textit{trap qubit}. Any other qubit is referred to as a \textit{dummy qubit}. Each trap qubit $v\in V_j$ will be $|+_{\theta_v}\rangle := (|0\rangle + e^{i\theta_v}|1\rangle )/\sqrt 2$, for some angle $\theta_v$ that is chosen uniformly at random from $\Theta = \{\frac{k\pi}{4}: k = 0,1,...,7 \}$. Each dummy qubit $v \in V\setminus V_j$ will be $|d_v\rangle$, where $d_v \in \{0,1 \}$ is chosen uniformly at random.  Then, the effect of the server applying its entangling gates is to flip each trap qubit to the orthogonal basis vector a number of times that corresponds to the sum (modulo 2) of the neighbouring dummies. This is a quantity that the client can compute. After constructing the graph state, the server measures its qubits and sends the outcome to the client. The trap qubit measurement basis that is specified by the client is $\left\{ |\pm_{\delta_v }\rangle \right\}$, for each trap $v\in V_j$, where $\delta_v = \theta_v + r_v \pi$, and $r_v \in \{ 0,1\}$ is chosen uniformly at random. The client compares the outcomes of the trap qubits to what is expected if all states and local operations are perfect, declaring the test round to be a failure if there is at least one trap measurement that is incorrect. A depiction of a graph state, a choice of $k$-colouring, and a choice of qubits for a test round is given in Figure \ref{fig:test_rounds}, for the case of a square graph.

\Figure[t!](topskip=0pt, botskip=0pt, midskip=0pt)[width=50mm]{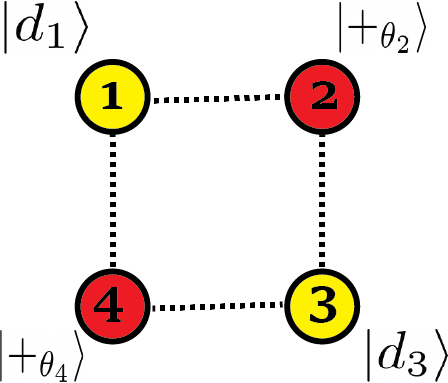}
{\textbf{Test rounds in the BQC protocol.} Here we choose an example where the computation is performed on a square graph. The client chooses a $k$-colouring: here $k=2$, and the two sets are coloured in yellow and red. In this case, the red vertices correspond to trap qubits, and the yellow vertices correspond to dummy qubits.\label{fig:test_rounds}}

\subsubsection{Error probability for a general graph}
We suppose that the client would like to know the outcome of a BQP calculation, which has corresponding graph $G=(V,E)$, and that the client has chosen a $k$-colouring $\{V_j \}_{j = 1}^k$. Then, given that the vector of fidelities at the time the server applies its operations is $\vec{F} = (F_1,...,F_{|V|})$, here we obtain a general form for the probability of error of the test round, $P_{G}(\vec{F})$. This is a generalisation of what can be found in  \cite{delft_eindhoven}, where BQC with two qubits is considered.

We firstly find the probability of error, \textit{given} that the client has chosen trap colour $V_{j}$. Call this $P_{V_{j}}$. The client thus chooses to send the trap qubits $v\in V_j$ as states $|+_{\theta_v} \rangle$. Then, at the time when the test round is carried out, the trap qubits corresponding to vertices $v \in V_j$ are each in a state $\rho_{v}$, where 
\begin{equation}
    \rho_{v} = F_{v} | +_{\theta_v} \rangle \langle +_{\theta_v} | + (1-F_{v}) | -_{\theta_v} \rangle \langle -_{\theta_v} | + (\text{o.d.1}),
    \label{eqn:qubit_state_trap}
\end{equation}
where we use (o.d.1) to write the off-diagonal elements (with respect to the basis $\{|+_{\theta_v}\rangle , |-_{\theta_v}\rangle  \}$). We don't write them out in full because these end up making no contribution to $P_{V_j}(\vec{F})$, as we will see later. Similarly, the dummy qubits $v \in V \setminus V_j$ will be in the state
\begin{equation}
    \rho_v = F_v |d_v \rangle \langle d_v)| + (1-F_v) | d_v \oplus 1 \rangle \langle d_v \oplus 1| + (\text{o.d.2}),
    \label{eqn:qubit_state_dummy}
\end{equation}
where here we use (o.d.2) to write the off-diagonal elements, this time with respect to the computational basis. The state of the server is then given by the tensor product of all of these states
\begin{equation}
    \rho_{\text{server}} = \bigotimes_{v \in V_j} \rho_v \bigotimes_{w \in W_j} \rho_w \bigotimes_{u \in V \setminus (V_j \cup W_j)} \rho_u, 
    \label{eqn:server_state_1}
\end{equation}
where we have defined $W_j \subset V\setminus V_j$ to be the set of all vertices that share an edge with a trap qubit. The server then proceeds with the next step of the BQC protocol, and applies $CZ$ gates to all pairs of qubits corresponding to edges in $E$, resulting in the state 
\begin{equation}
    \rho_{\text{server}}' = U \rho_{\text{server}} U^{\dag},
    \label{eqn:server_state_2}
\end{equation}
where $U : = \prod_{(w,v) \in E} CZ_{(w,v)}$. Recall that we are interested in the probability $P_{V_j}(\vec{F})$ that the this results in an error. In fact, is it simpler to find a form for the \textit{success probability} $Q_{V_j} (\vec{F}) = 1 - P_{V_j}(\vec{F})$. An error occurs when at least one of the trap qubit measurements does not match the result that would be obtained if all states were perfect. In particular, if everything were perfect, then the client would expect the measurement outcome corresponding to the trap qubit $v$ to be $r_v \oplus D_v$, where
\begin{equation}
    D_v =  \bigoplus_{w\in W_j : (v,w) \in E } d_w ,
\end{equation} 
i.e. the sum (modulo 2) of all the dummy variables $d_v$ that surround the trap qubit. The success probability is then given by
\begin{equation}
    Q_{V_j}(\vec{F}) = \text{Tr}_{V}\left( \left( \bigotimes_{v \in V_j} |(-1)^{r_v\oplus D_v}_{\delta_{v}} \rangle \langle (-1)^{r_v\oplus D_v}_{\delta_v } |\right) \rho_{\text{server}}' \right),
    \label{eqn:success_prob_def}
\end{equation}
where for convenience, we are using the notation $|(+1)_{\theta} \rangle  \equiv |+_{\theta} \rangle$ and $|(-1)_{\theta} \rangle  \equiv |-_{\theta} \rangle$. Rewriting $\ket{(-1)^{r_v+D_v}_{\delta_v }}=\ket{(-1)^{D_v}_{\theta_v}}$, and after examining Equations (\ref{eqn:server_state_1}) and (\ref{eqn:server_state_2}), we see that the qubit registers corresponding to vertices $v \in V \setminus (V_j \cup W_j)$ will make no contribution to this quantity, so that
\begin{equation}
    Q_{V_j}(\vec{F}) = \text{Tr}_{V_j} \text{Tr}_{ W_j} \left( \left( \bigotimes_{v \in V_j}|(-1)^{D_v}_{\theta_{v}} \rangle \langle (-1)^{D_v}_{\theta_v} | \right) \sigma_{\text{server}}' \right),
\end{equation}
with 
\begin{equation}
    \sigma_{\text{server}}' := \tilde{U} \sigma_{\text{server}} \tilde{U}^{\dag},
\end{equation}
where we have defined $\sigma_{\text{server}} := \bigotimes_{v \in V_j} \rho_v \bigotimes_{w \in W_j} \rho_w$ and $\tilde{U} := \prod_{(w,v) \in E_j} CZ_{(w,v)}$, and $E_j:= \{(v,w)\in E : v\in V_j \}$ to be the set of all edges between any element of $V_j$ and any other vertex. Recalling the states of our qubits as given in  (\ref{eqn:qubit_state_trap}) and (\ref{eqn:qubit_state_dummy}), and defining $F^{(0)} := F$, $F^{(1)} :=1-F$ to be used as a more concise way to write some of the terms, we can then write
\begin{align}
    \sigma_{\text{server}}  = & \bigotimes_{v \in V_j} \left( \sum_{x_v \in \{0,1 \} }  F_v^{(x_v)} | (-1)^{x_v}_{\theta_v} \rangle \langle (-1)^{x_v}_{\theta_v} |  \right) \cdot \notag\\ & \bigotimes_{w \in W_j} \left(\sum_{y_w \in \{0,1\}}  F_w^{(y_w)} | d_w+y_w \rangle \langle d_w+y_w | \right) \notag\\
        = \bigotimes_{v \in V_j} & \left( \sum_{x_v \in \{0,1 \} } F_v^{(x_v)} | (-1)^{x_v}_{\theta_v} \rangle \langle (-1)^{x_v}_{\theta_v} | \right) \cdot \notag\\  & \bigotimes \left( \sum_{\vec{y} \in \{0,1 \}^{|W_j|}} \prod_{w \in W_j} F_w^{(y_w)} |\vec{d}+\vec{y} \rangle \langle \vec{d}+\vec{y} |  \right),
    \label{eqn:server_state_3}
\end{align}
where in (\ref{eqn:server_state_3}) we have rewritten the sum to be over all length-$|W_j|$ binary strings $\vec{y} \equiv (y_w)_{w\in W_j} \in \{0,1 \}^{|W_j|}$. We have also stored the dummy variables in a vector $\vec{d}$, so that $(\vec{d}+\vec{y})_w=d_w+y_w$.  Again, we are not writing out the off-diagonal terms because these all disappear when we take the trace, and therefore make no contribution to the final expression. Applying the unitary operator $\tilde{U}$ yields
\begin{align*}
    \sigma_{\text{server}}' = \sum_{\vec{y} \in \{0,1 \}^{|W_j|}} & \prod_{w \in W_j} F_w^{(y_w)} | \vec{d}+\vec{y} \rangle \langle \vec{d}+\vec{y} | \cdot   \\ \bigotimes_{v \in V_j}  \Bigg( \sum_{x_v \in  \{0,1 \} } F_v^{(x_v)} & | (-1)^{x_v + s_v(\vec{y})+D_v}_{\theta_v} \rangle \langle (-1)^{x_v+s_v(\vec{y})+D_v}_{\theta_v} |  \Bigg) 
\end{align*}
where for a trap $v\in V_j$, we have defined 
\begin{equation}
    s_v(\vec{y}) := \sum_{w \in W_j : (w,v) \in E_j} y_w,
\end{equation}
which is the sum of the binary variables $y_w$ over all vertices neighbouring $v$. We can now start to trace out registers in order to find a final expression for $Q_{V_j}(\vec{F})$ in terms of the qubit fidelities. Taking the inner product $\langle (-1)^{D_v}_{\delta_v} | ... | (-1)^{D_v}_{\delta_v} \rangle $ for all trap qubits $v\in V_j$ and tracing out $W_j$ yields our final expression for the success probability as introduced in Equation (\ref{eqn:success_prob_def}),
\begin{equation}
    Q_{V_j}(\vec{F})=\sum_{\vec{y} \in \{0,1 \}^{|W_j|}}  \prod_{w \in W_j} F_w^{(y_w)}  \prod_{v \in V_j} F_v ^{(s_v(\vec{y}))} ,
    \label{eqn:success_prob_polynomial}
\end{equation}
This is a polynomial in the fidelities $\vec{F}=(F_1,...,F_{|V|})$, with a form that is completely determined by the graph structure and choice of trap colour $V_j$. The same thus holds for the error probability $P_{V_j}(\vec{F})$. In our model as given in Section \ref{sec:bqc_model}, it is further necessary to incorporate the fact that the first qubit to be sent is chosen at random. The probability of error is then effectively symmetrised over the $|V|$ possible starting qubits in the following way. Without loss of generality, letting the order in which the qubits are sent to be lexicographical, the probability of error is then
\begin{equation}
    P_{V_j}(\vec{F}) \rightarrow \tilde{P}_{V_{j}}(\vec{F}) := \frac{1}{|V|} \sum_j P_{V_j}(\sigma^j \vec{F}) ,
    \label{eqn:symmetrisation_poe}
\end{equation}
where $\sigma$ is the permutation that moves the vector elements one place to the left, i.e. $\sigma (F_1,...,F_{|V|} ) = (F_2,...,F_{|V|},F_1)$. To obtain the final probability of error, it remains to average over the choice of trap colour, recalling that this is chosen uniformly at random. This gives us a final expression for $P_G(\vec{F})$,
\begin{equation}
    P_G(\vec{F}) = \frac{1}{k}\sum_{j = 1}^k \tilde{P}_{V_{j}}(\vec{F}).
\end{equation}
\subsubsection{Error probability for a square graph}
An example of such a polynomial for the case of a square graph is as follows. Consider the $k$-colouring as in Figure \ref{fig:test_rounds}, with red as the choice of trap colour. Suppose that when the server applies its gates and measurements, the qubits have fidelities $\vec{F} = (F_1,F_2,F_3,F_4)$. Then, according to (\ref{eqn:success_prob_polynomial}), the success probability is given by 
\begin{gather*}
    Q_{\text{red}}(\vec{F}) = F_1 F_2 F_3 F_4 + F_1 (1-F_2) (1-F_3) (1-F_4) \\ + (1-F_1) (1-F_2) F_3 (1-F_4) + (1-F_1)F_2 (1-F_3)F_4,
\end{gather*}
and the error probability is then 
\begin{equation}
    P_{\text{red}}(\vec{F}) = 1-Q_{\text{red}}(\vec{F}).
\label{eqn:error_polynomial_square_graph}
\end{equation}
By symmetry of the square graph, the error probability $P_{\text{yellow}}(\vec{F})$ is obtained by exchanging the indices $1 \leftrightarrow 2, 3 \leftrightarrow 4$ in (\ref{eqn:error_polynomial_square_graph}).
In the case of the square graph, the symmetrisation maps $P_{\text{red}}(\vec{F})\rightarrow \tilde{P}_{\text{red}}(\vec{F})$, where
\begin{align*}
    \tilde{P}_{\text{red}}(\vec{F}) =  & \frac{1}{4}\big[P_{\text{red}}(F_1,F_2,F_3,F_4)+P_{\text{red}}(F_2,F_3,F_4,F_1) \\ & + P_{\text{red}}(F_3,F_4,F_1,F_2)+P_{\text{red}}(F_4,F_1,F_2,F_3)\big] .
\end{align*}
Note that the symmetries of the error functions $P_{\text{red}}$ and $P_{\text{yellow}}$ reflect the symmetries of the graph, i.e. they are symmetric under the interchange of $1\leftrightarrow 3$ or $2\leftrightarrow 4$. Then,
\begin{equation}
    \tilde{P}_{\text{red}}(\vec{F}) =  \frac{1}{2} \left(P_{\text{red}}(\vec{F}) + P_{\text{yellow}}(\vec{F}) \right) 
\end{equation}
The other error function $P_{\text{yellow}}$ maps to the same after the symmetrisation (\ref{eqn:symmetrisation_poe}), i.e. $\tilde{P}_{\text{yellow}}(\vec{F})=\tilde{P}_{\text{red}}(\vec{F})$. The probability of error, then, is given by
\begin{gather}
    P_{\text{square}}(\vec{F}) = \frac{1}{2}\left( \tilde{P}_{\text{yellow}}(\vec{F})+\tilde{P}_{\text{red}}(\vec{F}) \right) \\ = \frac{1}{2} \left(P_{\text{red}}(\vec{F}) + P_{\text{yellow}}(\vec{F}) \right). 
\end{gather}
This is the function that we use with (\ref{eqn:av_poe}) to calculate $p_{\text{av}}$ for our model, and compute the results for an example of a square graph in Section \ref{sec:bqc_results}.

\bibliographystyle{IEEEtran}
\bibliography{main}

\begin{thebibliography}{10}
\bibitem{github_repository}
Github repository to be made available.
\end{thebibliography}

\EOD 

\end{document}